\documentclass[11pt]{article}
\pdfoutput=1

\usepackage{jcappub}

\usepackage{color}
\usepackage{appendix}
\usepackage{latexsym,amsmath,amssymb,graphicx,booktabs}
\usepackage{epsfig,latexsym}
\usepackage{hyperref}
\numberwithin{equation}{section}
\usepackage{url}

\graphicspath{ {./Figures/} }

\definecolor{MyBlue}{rgb}{0.15,0.15,0.70}
\hypersetup{
colorlinks=true,
citecolor=MyBlue,
linkcolor=MyBlue,
urlcolor=MyBlue
}
\definecolor{lightgray}{gray}{0.9}

\def\newacronym#1#2#3{\gdef#1{#3 (#2)\gdef#1{#2}}}

\def\gw#1{gravitational-wave#1 (GW#1)\gdef\gw{GW}}
\def\grb#1{gamma ray burst#1 (GRB#1)\gdef\grb{GRB}}

\newacronym{\NR}{NR}{Numerical Relativity}
\newacronym{\GR}{GR}{General Relativity}

\def\bea{\begin{eqnarray}}
\def\eea{\end{eqnarray}}

\newcommand{\beq}{\begin{equation}}
\newcommand{\eeq}{\end{equation}}

\newcommand{\msun}{\mathrm{M}_{\odot}}
\newcommand{\Msun}{\mathrm{M}_{\odot}}

\def\bea{\begin{eqnarray}}
\def\eea{\end{eqnarray}}

\newacronym{\APPEC}{APPEC}{Astro-Particle Physics European Consortium}
\newacronym{\ILC}{ILC}{International Linear Collider}
\newacronym{\FALC}{FALC}{Funding Agencies for Large Colliders}
\newacronym{\ICFA}{ICFA}{International Committee for Future Accelerators}
\newacronym{\TMT}{TMT}{Thirty Meter Telescope}

\title{Science Case for the Einstein Telescope}
\author[a]{Michele~Maggiore,}
\author[b]{Chris~Van~Den~Broeck,}
\author[c,d,e]{Nicola~Bartolo,}
\author[a]{Enis~Belgacem,}
\author[c,d]{Daniele~Bertacca,}
\author[f]{Marie~Anne~Bizouard,}
\author[g,h]{Marica~Branchesi,}
\author[i,j]{Sebastien Clesse,}
\author[a]{Stefano~Foffa,}
\author[k]{Juan~Garc\'ia-Bellido,}
\author[g,h]{Stefan~Grimm,}
\author[g,h]{Jan~Harms,}
\author[l]{Tanja~Hinderer,}
\author[c,d,e,g]{Sabino~Matarrese,}
\author[m]{Cristiano~Palomba,}
\author[c,d]{Marco~Peloso,}
\author[d]{Angelo~Ricciardone,}
\author[n]{and Mairi~Sakellariadou}

\affiliation[a]{D\'epartement de Physique Th\'eorique and Center for Astroparticle Physics,\\
Universit\'e de Gen\`eve, 24 quai Ansermet, CH--1211 Gen\`eve 4, Switzerland}
\affiliation[b]{Department of Physics, Utrecht University
Princetonplein 1, 3584 CC Utrecht  and  
Nikhef, National Institute for Subatomic Physics,
Science Park 105, 1098 XG Amsterdam,
The Netherlands}
\affiliation[c] {Dipartimento di Fisica e Astronomia Galileo Galilei, Universit\`a di Padova, 35131 Padova, Italy}
\affiliation[d] {INFN, Sezione di Padova, via Marzolo 8, I-35131, Padova, Italy}
\affiliation[e]{INAF, Osservatorio Astronomico di Padova,  vicolo dell'Osservatorio 5, I-35122 Padova, Italy}
\affiliation[f]{Artemis, Universit\'e C\^ote d'Azur, Observatoire C\^ote d'Azur, CNRS, CS 34229, F-06304 Nice Cedex 4, France}
\affiliation[g]{Gran Sasso Science Institute (GSSI), I-67100 L'Aquila,
Italy}
\affiliation[h]{INFN, Laboratori Nazionali del Gran Sasso, I-67100 Assergi,
Italy}
\affiliation[i]{Cosmology, Universe and Relativity at Louvain (CURL), Institut de Recherche en Mathematique et Physique (IRMP), Louvain University, 2 Chemin du Cyclotron, 1348 Louvain-la-Neuve, Belgium}
\affiliation[j]{Namur Institute of Complex Systems (naXys), Department of Mathematics,
University of Namur, Rempart de la Vierge 8, 5000 Namur, Belgium}
\affiliation[k]{Instituto de F\'isica Te\'orica UAM-CSIC, Universidad Auton\'oma de Madrid,
Cantoblanco, 28049 Madrid, Spain}
\affiliation[l]{Delta Institute for Theoretical Physics, Science Park 904, 1090 GL Amsterdam, The Netherlands, and
GRAPPA, Anton Pannekoek Institute for Astronomy and Institute of High-Energy Physics, University of Amsterdam, Science Park 904, 1098 XH Amsterdam, The Netherlands
}
\affiliation[m]{INFN, Sezione di Roma, I-00185 Roma, Italy}
\affiliation[n]{Theoretical Particle Physics and Cosmology Group, Physics Department,
King's College London, University of London, Strand, London WC2R 2LS, United Kingdom
}

\abstract{The Einstein Telescope (ET), a proposed European ground-based gravitational-wave detector of  third-generation,  is an evolution of  second-generation detectors such as Advanced LIGO, Advanced Virgo, and KAGRA which could be operating in the mid 2030s. ET will explore the universe with gravitational waves up to cosmological distances. 
We discuss its main scientific objectives and its potential for discoveries in astrophysics, cosmology and fundamental physics.\footnote{Prepared for submission to the ESFRI Roadmap, on behalf of the ET
steering committee.}}

\begin{document}
\maketitle
\flushbottom

\section{Introduction}\label{chap:ScienceCase}

The gravitational-wave (GW) detectors of second generation (2G), Advanced LIGO and Advanced Virgo, have truly opened a new window on the Universe. The first direct detection of GWs from a binary black hole (BH) coalescence, in September~2015~\cite{Abbott:2016blz},  was a historic moment, and the culmination of decades of efforts from a large community. Another historic moment was the first detection of a neutron star (NS) binary coalescence, together with the simultaneous detection of the associated gamma-ray burst, and the subsequent observation of the electromagnetic counterpart in all bands of the electromagnetic 
spectrum~\cite{TheLIGOScientific:2017qsa,Goldstein:2017mmi,Savchenko:2017ffs,Monitor:2017mdv,GBM:2017lvd,Coulter:2017,Troja:2017,Hallinan:2017}. A number of additional detections have taken place since,  to the extent that, at the current level of sensitivity of 2G experiments, BH-BH detections are taking place on a weekly basis.
Many remarkable results in astrophysics and in fundamental physics have already been obtained thanks to these first detections. To mention only a few highlights,  the observation of the NS-NS binary coalescence GW170817 solved the long-standing problem of the origin of (at least some) short gamma ray bursts~\cite{TheLIGOScientific:2017qsa,Mooley:2018,Ghirlanda:2019}; the multi-band observations of the associated kilonova revealed that NS-NS mergers are a site for  the  formation of some of the heaviest elements through r-process nucleosynthesis~\cite{Pian:2017,Smartt:2017,Kasen:2017,Watson:2019}; the observation of tens of BH-BH coalescences has  revealed a previously unknown population of stellar-mass BHs, much heavier than those detected through the observation of X-ray binaries~\cite{LIGOScientific:2018mvr}, and has shown that BH-BH binaries exist, and coalesce within a Hubble time at a detectable rate.  Concerning  fundamental physics, cosmology and General Relativity (GR), the observation of the GWs and the gamma-ray burst from the NS-NS binary GW170817 proved that the speed of GWs is the same as the speed of light to about a part in $10^{15}$~\cite{Monitor:2017mdv}; the GW signal, together with the electromagnetic determination of the redshift of the source, provided the first measurement of the Hubble constant with GWs~\cite{Abbott:2017xzu};  the tail of the waveform of the first observed event, GW150914, showed  oscillations consistent with the prediction from General Relativity  for the quasi-normal modes of the final BH~\cite{TheLIGOScientific:2016src}; several  possible deviations from GR (graviton mass, post-Newtonian coefficients, modified dispersion relations, etc.) could be tested and bounded~\cite{TheLIGOScientific:2016src,Abbott:2018lct,LIGOScientific:2019fpa}.

Extraordinary as they are, these results can however be considered only as a first step toward our exploration of the Universe with GWs.
Third-generation (3G) GW detectors, like the Einstein Telescope (ET)~\cite{Punturo:2010zz}, will bring the  gravitational wave astronomy revolution to a full realisation. Thanks to an order of magnitude better sensitivity and a wider accessible frequency band with respect to second-generation detectors, 3G detectors will allow us to address a huge number of key issues related to  astrophysics, fundamental physics  and cosmology. 

\begin{figure}[t]
\centering
\includegraphics[width=0.45\textwidth]{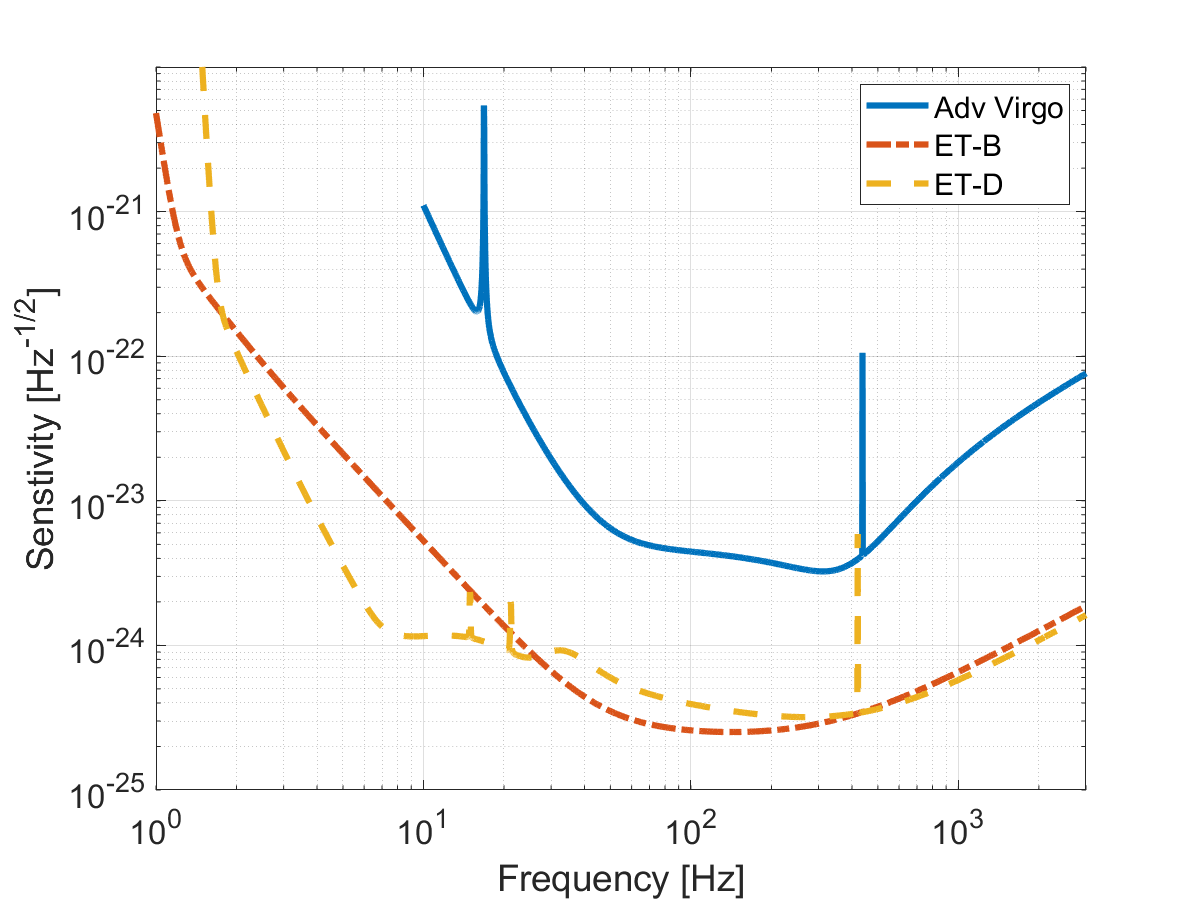} 
\includegraphics[width=0.535\textwidth]{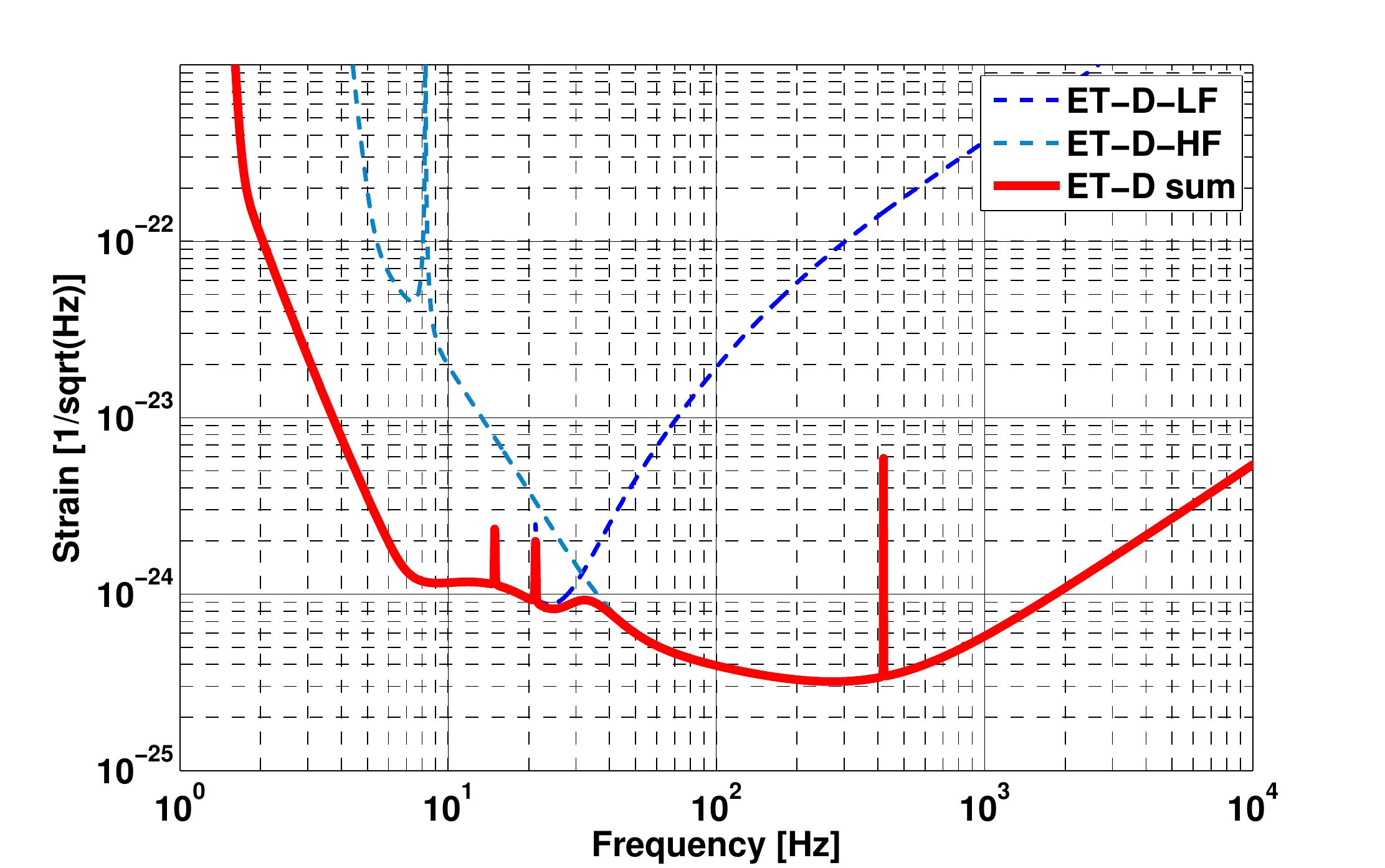} 
\caption{Left panel: the sensitivity of the ET-B (red, dot-dashed) and ET-D (yellow, dashed) configurations, compared to the official target sensitivity of advanced  Virgo (blue solid line).
Right panel: the separate contributions from the LF and HF instruments to the sensitivity of ET-D.}
\label{fig:ETsens}
\end{figure}

A detailed and updated discussion of the ET technical design will be presented in~\cite{ETCDR2019}, while here we will focus on the science that can be made with ET. Let us nevertheless briefly summarize some of the main features  that constitute the core of the ET design. ET will be a ground-based GW interferometer, that builds on the experience gained with 2G detectors. The arm length $L$ will be increased to 10~km, compared to 3~km for Virgo and 4~km for LIGO. This will reduce all displacement noises.\footnote{In a first approximation, coating thermal noise decreases as $1/L$, but there are also additional gains; for instance,  since also the
beam diameter increases on the mirrors, one gets  better averaging of
thermal noise over the beam area~\cite{Evans:2016mbw}.} To reduce gravity gradient noise and seismic noise, and therefore extend significantly the sensitivity toward low frequencies, ET will be built a few hundred meters underground.  Two  candidate sites remain and are currently under investigation: one in Sardinia, near the former Sos Enattos mine, and one  at the three-border region of Belgium-Germany-Netherlands. 
ET will have  a triangular shape,  corresponding to three nested interferometers. This will provide a more isotropic antenna pattern without blind spots and the possibility of fully resolving both GW polarizations, and  will also allow for redundancy, including the possibility of constructing  null streams from the combined outputs.
To reduce shot noise and therefore improve the sensitivity in the high frequency regime,  the laser power  at the interferometer input will be increased, leading to high optical power in the cavity of the order of  several MW, while (in the simpler design, see below) brownian  noise in the mirror coating will  be reduced by cooling the test masses to cryogenic temperatures, down to 20~K. These improvements, together with several others (optimisation of signal recycling, frequency-dependent light squeezing, increase of the beam size, heavier mirrors, etc.) were at the core of the so-called ET-B sensitivity curve~\cite{Hild:2008ng}, shown in the  left panel of Fig.~\ref{fig:ETsens} (where, for comparison, we also show the target sensitivity of advanced Virgo). However, this curve neglects the fact that, in practice, high circulating power is difficult to reconcile with cryogenic test masses. Indeed, with several MW of circulating light power, even with state-of-the-art coating technology, providing  a residual absorption  of just one ppm in the mirror  coatings, we still have several Watt of absorbed power in the coatings.
This requires to increase the   thickness of the suspension fibers in order  to remove the generated heat which, in turn,  would spoil the performance of the suspension system and therefore the sensitivity at low-frequencies.
These considerations have led to a `xylophone' concept, in which the detector is actually composed of two different instruments, one optimized for low frequencies  (LF) and one for high frequencies (HF); the LF interferometer has low power (since laser power is only needed to beat down the shot noise in the high frequency range) and cryogenic mirrors, while the HF instrument has high power and mirrors at room temperature. This has lead to the ET-C sensitivity curve and, after some further refinement of some noise models, to the
ET-D sensitivity curve \cite{Hild:2010id}, again shown in  the left panel Fig.~\ref{fig:ETsens}, while the right panel shows also the separate contributions from the LF and HF interferometers. The ET-D sensitivity curve will be the baseline sensitivity that we will adopt in this paper. Occasionally, in some plots we will also compare with the results for the ET-B sensitivity, to appreciate the dependence of the results on the sensitivity curve.

\begin{figure}[t]
 \centering
 \includegraphics[width=0.4\textwidth]{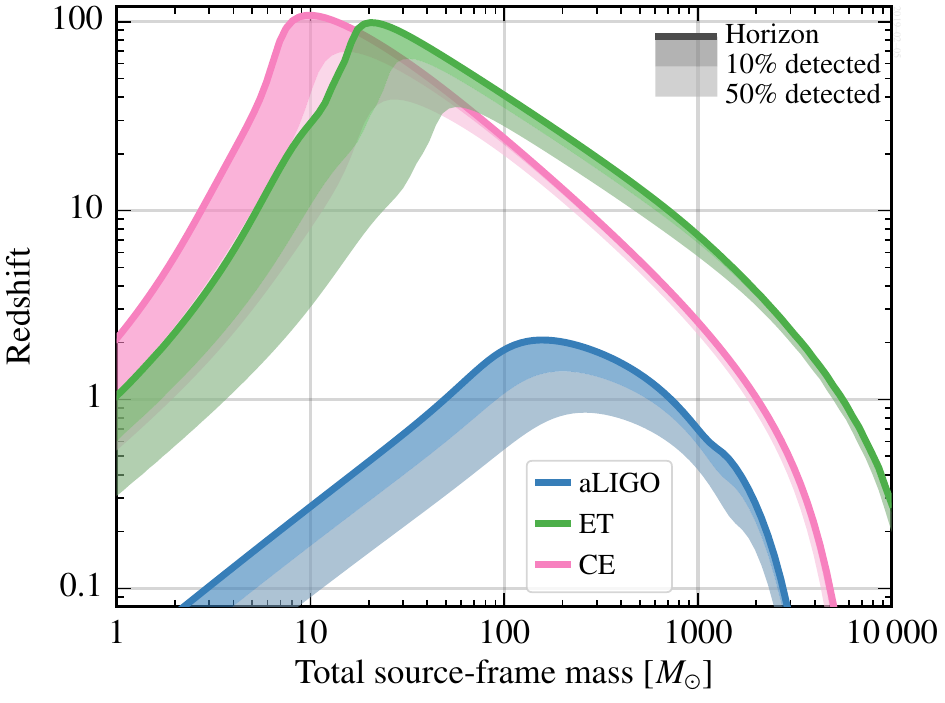}\qquad\qquad
 \hspace{-0.09\textwidth}
  \includegraphics[width=0.4\textwidth]{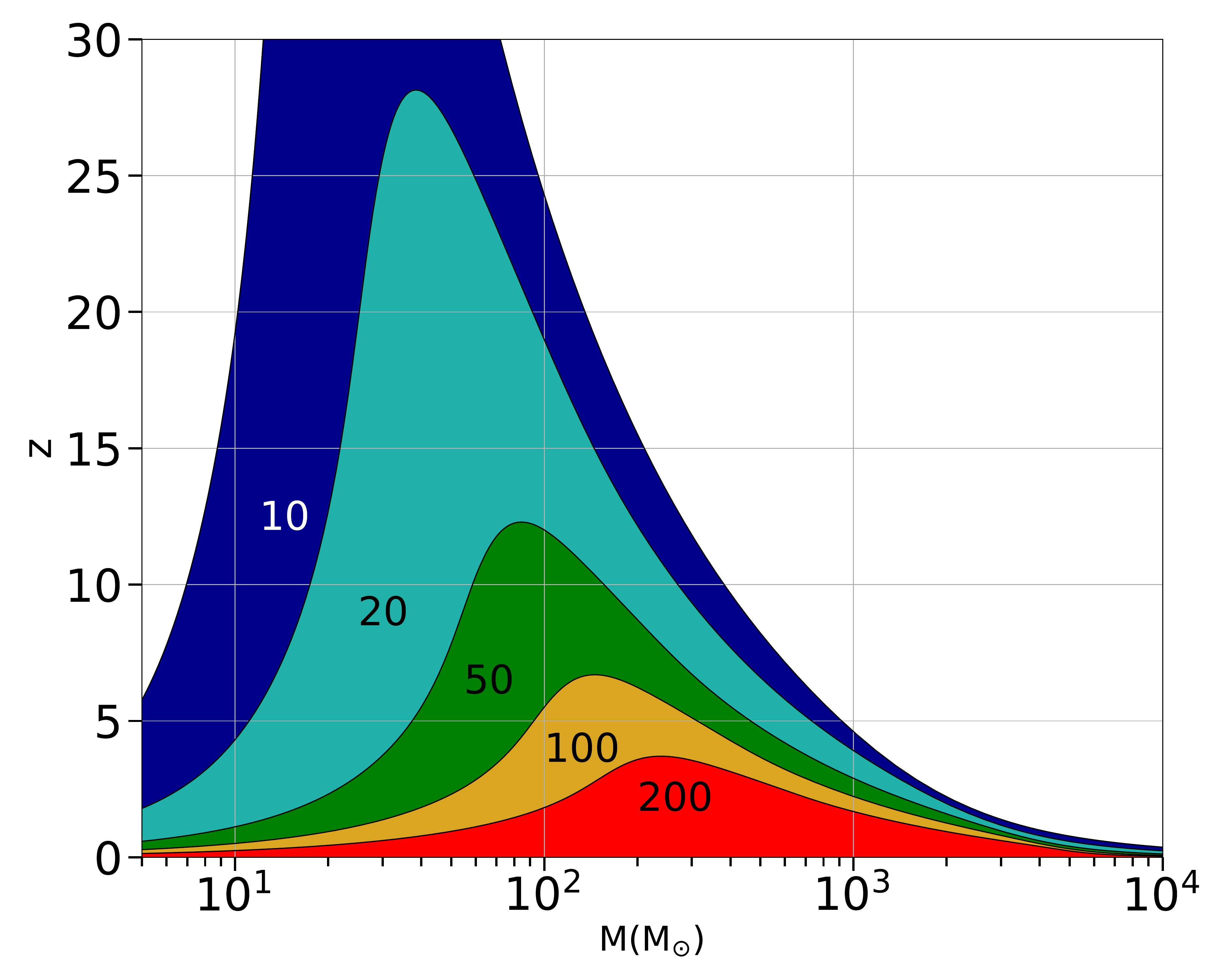}
 \caption{Left: astrophysical reach  for equal-mass, nonspinning binaries  for Advanced LIGO, Einstein Telescope and Cosmic Explorer (from ref.~\cite{Sathyaprakash:2019nnu,3GScienceBook}). Right:
lines of constant signal-to-noise ratio in the  (total mass, redshift)  plane, for a network of one ET and two CE detectors. The curves shown assume equal-mass binary components (figure courtesy by M. Colpi and A. Mangiagli).}
\label{fig:gw_horizons}
\end{figure}

An example of   the extraordinary potential of 3G detectors is provided  by Fig.~\ref{fig:gw_horizons}. The left panel shows the detector reach, in term of cosmological redshift, as a function of the total mass of a coalescing binary. We see that the coalescence of  compact binaries with total mass  $(20-100)~\Msun$, as typical of BH-BH or BH-NS binaries, will be visible by ET up to redshift $z\sim 20$ and higher, probing the dark era of the Universe preceding the birth of the first stars. (In particular, BH-BH mergers seen at such distances would necessarily have a primordial origin.)
By comparison, in the catalog of  detections from the O1 and O2 Advanced LIGO/Virgo runs, the farthest BH-BH event is at $z\simeq 0.5$ and, at final target sensitivity, 2G detectors should reach $z\simeq 1$. The range of BH masses accessible will also greatly increase; as we see from  Fig.~\ref{fig:gw_horizons}, ET will be able to detect BHs with masses up to several times $10^3~\Msun$, out to $z\sim 1-5$.

For NS-NS binaries, whose  total mass is around $3\,\Msun$,  ET will reach $z\simeq 2-3$; by comparison, the NS-NS binary GW170817 was at $z\simeq 0.01$ and, at final target sensitivity, 2G detectors should reach $z\simeq 0.2$.

The corresponding detection rates will be impressive, of order $10^5-10^6$ BH-BH and $7\times 10^4$ NS-NS coalescences per year for a single detector such as ET~\cite{Regimbau:2012ir,Regimbau:2014uia,Belgacem:2019tbw}; depending on the network of electromagnetic facilities operating at the time of 3G detectors, over a few years one might collect $O(10^2-10^3)$ NS-NS GW events with observed electromagnetic counterpart~\cite{Belgacem:2019tbw}. The signal-to-noise ratio of many of these events will be huge.
The right panel in Fig.~\ref{fig:gw_horizons} shows that, in a network of three 3G detectors,  a BH-BH binary coalescence with total mass $(50-100) \Msun$ at $z=10$  could be seen  with signal-to-noise ratio of order 50, and the signal-to-noise ratio of some events, still at cosmological distances, will  even be of order 100-200, which will allow us to determine the shape of the gravitational wave with exquisite precision.
{\em The combination of distances and masses explored, sheer number of detections, and detections with very high signal-to-noise ratio will provide a wealth of data that have the potential of triggering revolutions in astrophysics, cosmology and fundamental physics.}

Beside coalescing binary systems, ET will be able to detect several other kinds of signals, such as stochastic backgrounds of GWs, signals from isolated pulsars, or supernovae, with a sensitivity that improves by orders of magnitude compared to 2G detectors.

As we shall see, many of the possible achievements of ET, and other planned 3G detectors like Cosmic Explorer in the U.S., are only possible through gravitational waves. For others, GW detectors are complementary to  facilities exploiting electromagnetic radiation  or other messengers, such as neutrinos and cosmic rays. Combined observations through GWs, electromagnetic signals, neutrinos and/or cosmic rays, will give us a multi-messenger and more comprehensive picture of many energetic phenomena of the Universe.
Schematically, we can identify the following main items as part of the ET science case:

\vspace{2mm}

\begin{itemize}
\item Astrophysics 
  \begin{itemize}
   \item Black hole properties: origin (stellar vs. primordial), evolution, demography.
   \item Neutron star properties: interior structure (QCD at ultra-high densities, exotic states of matter), demography.
   \item Multi-messenger astronomy: nucleosynthesis, physics of jets, 
role of neutrinos.
   \item Detection of new astrophysical sources of GWs: core collapse supernovae, isolated neutron stars, stochastic background of astrophysical origin.
  \end{itemize}
\item Fundamental physics and cosmology
  \begin{itemize}
   \item The  nature of compact objects: near-horizon physics, tests of no-hair theorem, exotic compact objects.
  \item Dark matter: primordial BHs, axion clouds, dark matter accreting on compact objects.
  \item Dark energy and modifications of  gravity on cosmological scales.
   \item Stochastic backgrounds of cosmological origin and connections with high-energy physics (inflation, phase transitions, cosmic strings, ...)
  \end{itemize}
\end{itemize}

 \vspace{2mm}\noindent
It should  be stressed, however, that many questions cross the borders between  domains outlined above. For instance, understanding whether the BHs observed by GW detectors are of stellar or primordial origin obviously has an astrophysical interest, but a primordial origin would have deep consequences on  our understanding of early Universe physics, inflation, etc., subjects that belong to the domain of cosmology and of fundamental physics. As another example, determining the equation of state in the core of neutron stars is of great importance both in astrophysics and for understanding the theory of strong interactions, QCD, in the regime of ultra-high  density, where phase transitions can take place.
 
In the following sections we briefly discuss some of the science that ET will be able to address. We will conclude with a summary of the Key Science Questions in Section~\ref{sect:ScienceCaseKeyQuestions},
underlying in particular what science a single ET observatory can achieve.
A more detailed discussion of the science of 3G detectors will be presented in~\cite{3GScienceBook}.

\section{Astrophysics}\label{sec:astrophy}

\subsection{Black hole binaries}\label{sect:Blackholebinaries}

\begin{figure}[t]
\centering
{\includegraphics[width=0.5\textwidth]{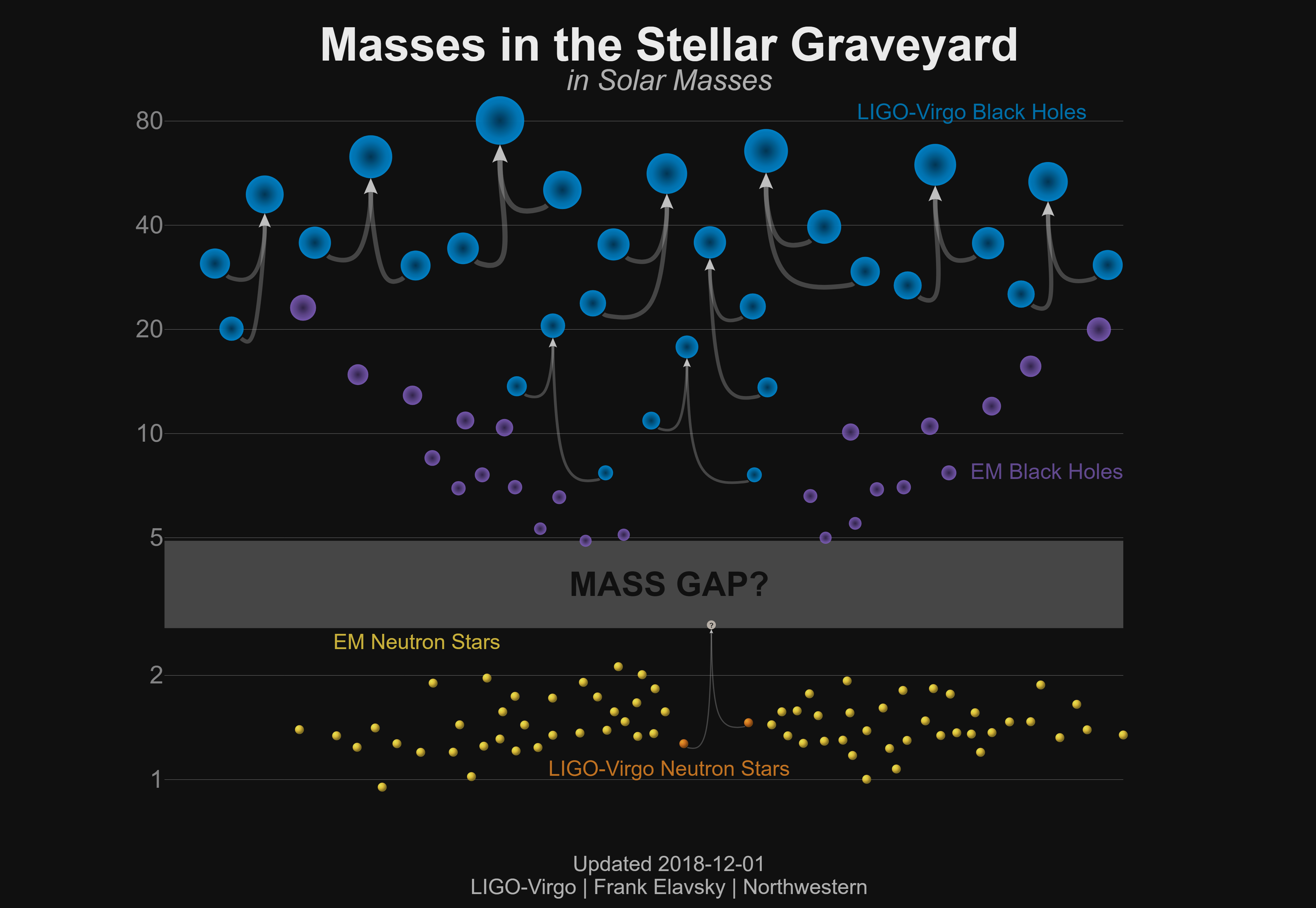}}
\caption{Masses of the LIGO/Virgo BHs detected during the O1 and O2 runs (light blue), of BHs discovered through X-ray binaries (purple),
NSs known electromagnetically (yellow) and the two initial  NS in the binary GW170817 (orange). Image taken from \url{https://media.ligo.northwestern.edu}. 
\label{fig:Masses}
}
 \end{figure}

Observationally, BHs have been first identified through X-ray binaries - binary systems in which a BH accretes matter from a companion star. The remarkable GW detections of Advanced LIGO/Virgo in the O1 and O2 runs have then revealed a whole new population of stellar-mass binary BHs with much higher masses, see Fig.~\ref{fig:Masses}.

With BH-BH and BH-NS coalescence, 3G detectors will explore the Universe to extraordinary depths, providing  an unobstructed view on the earliest universe up to the dark ages, and generally with higher SNR signals compared to current EM observations from these early times.\footnote{To understand how this is possible, it is useful to recall that a BH-BH coalescence such as the first  detected event, GW150914, converted  into GWs an energy of  $3 \Msun c^2$  in just the last few milliseconds of the coalescence. The peak luminosity of the event,  $3.6\times 10^{56}\, {\rm erg/s}$, or  $200\, \Msun c^2/{\rm s}$, was an order of magnitude  larger than the estimated combined electromagnetic luminosity of all star and galaxies in the observable Universe!} {\em ET will uncover the  full population of coalescing stellar and intermediate mass BHs in the Universe,
over the whole epoch since the end of the cosmological dark ages.}
This will allow ET to answer several key questions about the origin and evolution of BH-BH systems. In particular:

\vspace{1mm}\noindent
(1) The ET  observations of BH-BH binaries  across the whole epoch of star formation would contain evidence,  accessible in no other way, of the cosmic history of stellar evolution, including the earliest populations of stars formed in the Universe.   
Since it is expected that compact-object formation is a metallicity-dependent process, that information will be complementary to improving electromagnetic constraints on the star-formation history of the Universe.  The fact that ET  will observe BH-BH mergers beyond the reionization epoch, at $z \ge 6$, will enable the determination of features such as the masses of the first metal-poor progenitor stars, and the relation between star metallicity and BH masses. 

\vspace{1mm}\noindent
(2) GW observations of BH-BH binaries probe the physics of BH formation in situations which lead to mergers.  ET will provide some events with extraordinarily well-measured properties, alongside large samples of mergers from which statistical population characteristics can be extracted.   
This will allow us to understand how binary evolution results in BH-BH mergers, and will give information on several aspects of the dynamics of binaries, such as the impact of the common envelope phase in the progenitor binaries, or the role of the dynamics of star clusters and galactic nuclei in producing close binaries of compact objects. 
   
\vspace{1mm}\noindent
(3) Thanks to the fact that the  reach of ET for BH-BH systems is well beyond the peak of the star formation  at $z\sim 2$, by comparing the redshift dependence of the BH-BH merger rate with the cosmic star formation rate it will be possible  to disentangle the contribution of BHs of stellar origin from that of  possible BHs of  primordial origin  (whose merger rate is not expected to be correlated with the star formation density). Moreover, BH-BH systems of stellar origin are expected to form in galaxies, while primordial systems should trace the distribution of dark matter rather than that of baryons. The huge number of detections in ET will allow to perform cross-correlations between the detected GW events and large-scale structures, providing another clue to the origin of the observed BHs.
{\em Showing that at least a fraction of the observed BHs are of primordial origin would be a discovery of fundamental importance not only in astrophysics but also from the point of view of fundamental physics. }Primordial BHs of mass around a solar mass could have formed at the QCD quark-hadron transition via gravitational collapse of large curvature fluctuations generated during the 
last stages of inflation~\cite{Carr:2019hud}. This could explain not only the present abundance of dark matter but also the baryon asymmetry of the universe~\cite{Garcia-Bellido:2019vlf}. The large number of binary BH detections with ET, as a function of mass and redshift, would also allow us to obtain detailed characterization of the population of primordial BHs, their mass and spin distributions and their formation times as a function of mass, as well as their clustering properties, as inferred from their merger rates and the induced stochastic GW background~\cite{Clesse:2016vqa,Clesse:2016ajp,Garcia-Bellido:2017fdg}.

\vspace{1mm}\noindent
(4) The discovery of luminous quasars at redshift as large as $z\sim 7$ suggests that,  at  $z>7$, there should be a population of `seed' BHs, with masses in the range  $(10^2-10^5) \msun$, from which these supermassive BHs have grown through gas accretion.   Furthermore, possible seed BHs that were in an environment such that further growth by gas accretion was suppressed could  be present at smaller redshifts, with masses close to their initial values.
As we see from the left panel of Fig.~\ref{fig:gw_horizons}, ET has the sensitivity necessary to detect  BH binary systems containing  a  BH with mass between  $O(10^2)\,   \Msun$ and  a few times $10^3\,\Msun$, up to large redshifts.  ET could therefore detect these seed BHs, providing crucial missing links in the formation and evolution of structures in the early Universe, and  unraveling the possible connection between stellar-mass black holes and supermassive black-hole in the center of the galaxies.

\subsection{Neutron stars} 

\begin{figure}[t]
 \centering
 \includegraphics[width=0.40\textwidth]{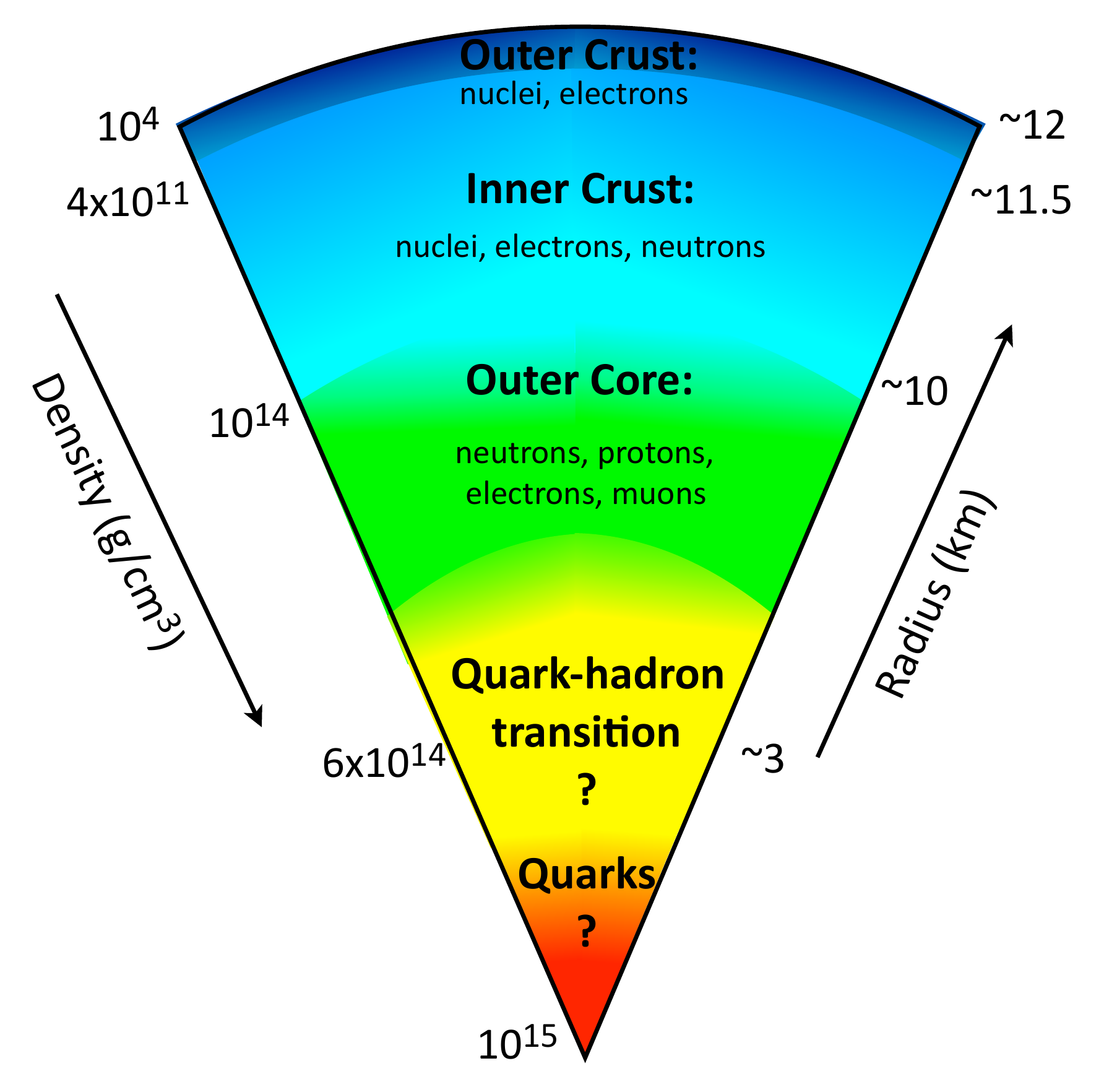}\qquad\qquad
  \includegraphics[width=0.48\textwidth]{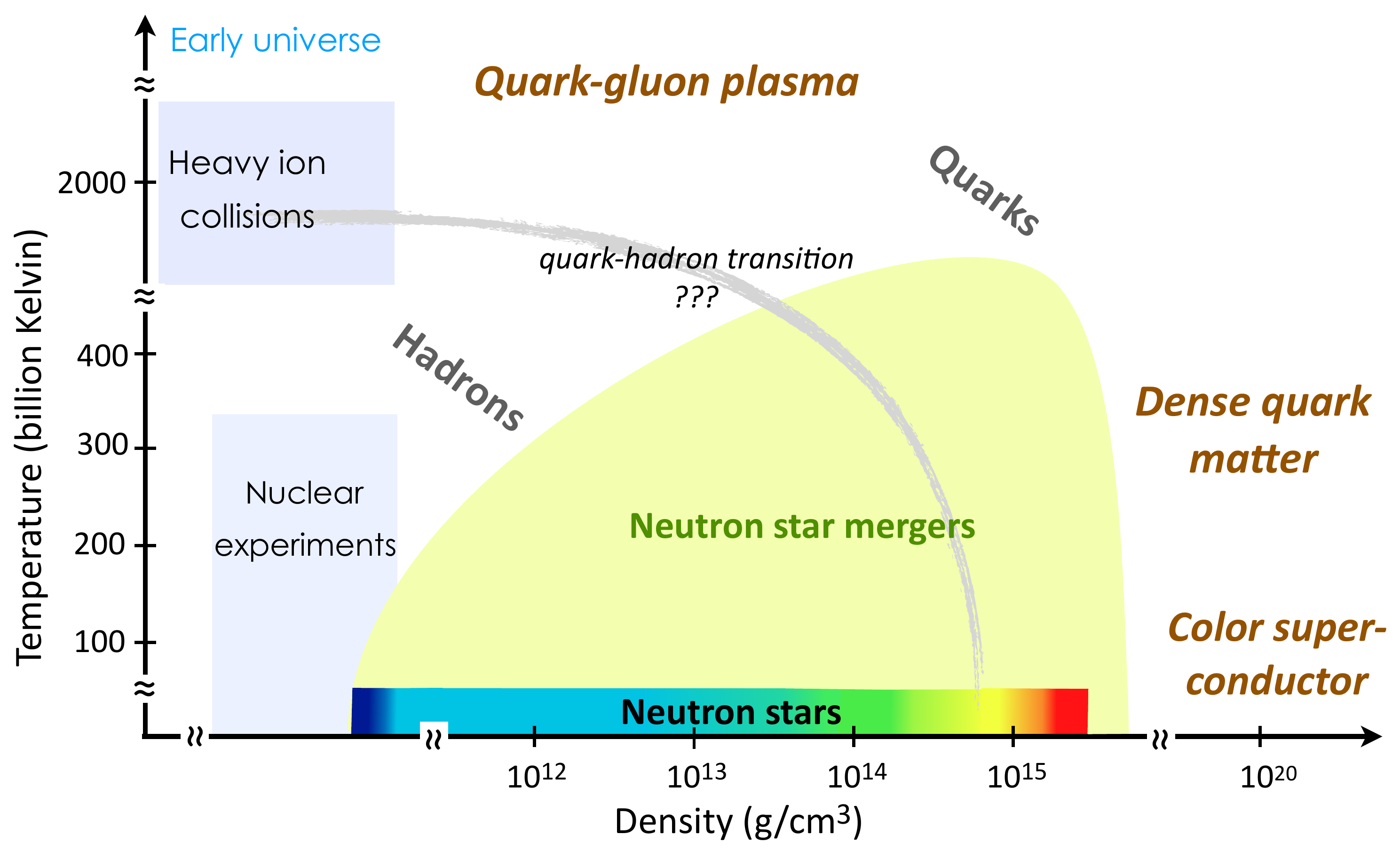}
 \caption{Left: Conjectured interior structure of a neutron star.  Right:
Matter encountered in neutron stars and binary mergers explores a large part of the QCD phase diagram in regimes that are inaccessible to terrestrial collider experiments.}
\label{fig:NSs}
\end{figure}

Neutron stars (NSs) are extraordinary laboratories for studying the fundamental properties of subatomic matter under conditions far from the realm accessible to experiments and first-principles theoretical calculations. In NSs, intense gravity compresses matter to several times the density of an atomic nucleus. Predicting the composition of such matter and the multi-body interactions providing sufficient pressure to prevent utter collapse to a BH requires large extrapolations from known physics and has been a longstanding scientific frontier. The left panel in Fig.~\ref{fig:NSs} illustrates the conjectured interior structure of NSs, spanning a vast range in density. Near the surface of a NS, neutron-rich nuclei and free electrons dominate, while at higher densities towards the interior the nuclear structure dissolves into primarily a uniform liquid of neutrons. In the cores of NSs yet more exotic states of matter may emerge, such as condensates of particles containing strange quarks. In addition, the quark substructure of the hadrons is expected to become important at densities a few times nuclear density, where states of matter comprising de-confined quarks may appear~\cite{Baym:2017whm}. The properties and parameter space of the transition between nuclear and quark matter and the states of quark matter encountered at low temperature, in the interaction-dominated ground state of dense matter, have been key questions in subatomic physics. Neutron stars thus provide a unique window onto the behavior of QCD, the fundamental theory of strong interactions, in a regime complementary to the higher temperatures and lower baryon densities accessible in collider experiments that probe the quark-gluon plasma. 

A complete understanding of the entire QCD phase diagram requires a synergy of the complementary insights gained from the terrestrial heavy-ion collision experiments and the astrophysical collisions of two NSs at close to the speed of light driven by dynamical gravity. In such NS mergers, matter encounters yet higher densities and temperatures than in individual NSs, thus providing a window onto a completely unexplored regime of subatomic physics. The right panel in Fig.~\ref{fig:NSs} shows the regimes in the QCD phase diagram relevant to NS binaries, which occupy a large swath of unexplored parameter space and a complex nonperturbative regime between our current understanding of matter at low baryon densities from nuclear physics at low temperature and heavy-ion collisions at higher temperatures, and the extremely high-densities where quark-matter calculations are valid. 

The fundamental properties of NS matter give rise to characteristic imprints in the GW signals from NS binaries or individual asymmetric NSs, making GWs unique probes of subatomic physics in unexplored regimes. A 3G GW detector with a high sensitivity and large frequency bandwidth such as ET will be critical to shed light on important fundamental physics questions, by precisely determining the properties of cold, dense matter in NSs as well as the new physics encountered during a binary NS merger.

\subsubsection{Coalescing neutron star binaries} 
With 2G detectors, the first observed NS-NS coalescence
GW170817 demonstrated that useful limits on the NSs' tidal deformability, a characteristic parameter that depends on the properties of matter in their interiors, could be extracted from the inspiral part of the GW signal~\cite{TheLIGOScientific:2017qsa}. However, despite the proximity of GW170817, the inferred constraints on the equation of state of NS matter ~\cite{Abbott:2018exr}
are too weak to discriminate between realistic models, nor do they offer 
new insights about phase transitions~\cite{Hebeler:2010jx,Gandolfi:2011xu,Tews:2018kmu}.

To determine in detail the nature of matter and interactions in NS interiors requires measuring tidal deformability with an order of magnitude higher accuracy. In addition, such high-fidelity measurements must be obtained for a population of NSs spanning a wide range of masses to map out the parameter dependencies and identify potential signatures of phase transitions. Both can be achieved with ET, which will detect a huge number of NS-NS coalescences per year as quantified above, and will observe their signal with order of magnitudes higher accuracy. As an example, an event like the first observed coalescing NS-NS binary, GW170817, would be seen at ET with a signal-to-noise ratio larger than the one of the event actually observed 
in LIGO/Virgo by a factor ${\cal O}(50)$, resulting in an overall signal-to-noise ratio as large 
as $\sim 1700$ (in the ET-D configuration).
This sensitivity could even allow us to access magnetic and rotational tidal Love numbers of the component neutron stars~\cite{Jimenez-Forteza:2018buh}.

The high-accuracy measurements performed with a 3G detector will also enable us to discern subdominant signatures of matter in GWs from binary inspirals that are inaccessible with 2G instruments. Such effects encode key information on details of the fundamental physics in NS interiors that are not probed directly with the tidal deformability. For instance, detecting GW signatures from the tidal excitation of a NS's internal oscillation modes during an inspiral, as will likely be possible with ET, would provide an unprecedented spectroscopic view of NS interiors and reveal exquisite details about the existence and nature of phase transitions to new states of matter.

A further unique capability of 3G detector such as ET with a high sensitivity extending to frequencies in the kiloHertz range is to open the major new scientific discovery space of matter phenomena beyond the inspiral. Witnessing the tidal disruption of a NS by a BH for a variety of systems will yield further insights into the properties of NS matter under extremes of gravity, and tracking the violent collision of two NSs and its aftermath will provide an exceptional window onto fundamental properties of matter in a completely unexplored regime, at higher temperatures and yet greater densities than encountered in individual NSs. 

\begin{figure}[t]
\center{\includegraphics[width=0.6\textwidth]{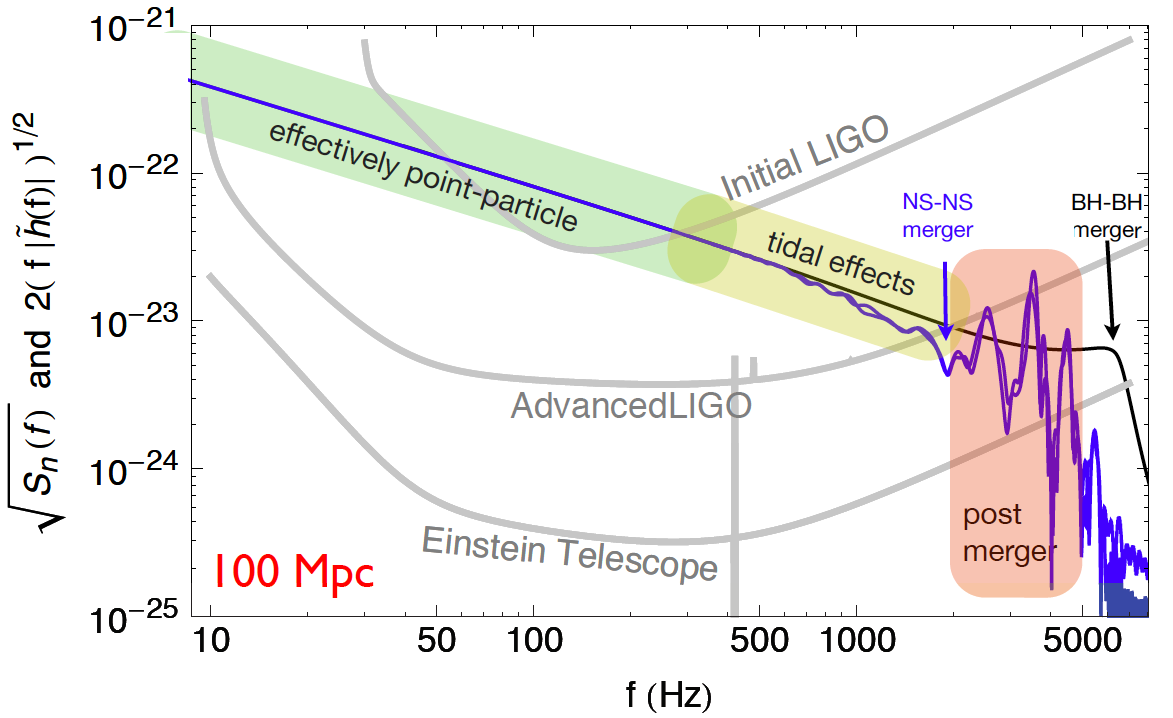}}
\vspace*{-0.4cm}
\caption{Gravitational wave signal from a NS-NS merger at a distance 100~Mpc, as it sweeps
  across the detector-accessible frequency range. From~\cite{Maggiore:2018zz} 
(figure courtesy of Jocelyn Read, based on results presented in \cite{Read:2013zra}).
}
\label{fig:merger_phases}
\end{figure}

The outcome of a binary NS merger strongly depends on the parameters. It is either a short-lived hypermassive NS that is temporarily stabilized by rotational effects yet ultimately collapses to a BH, or a BH that forms immediately upon merger, or a temporary supra-massive NS that settled to a NS remnant. The emitted GWs are distinct for the different scenarios and contain copious information on the complex microphysics. To fully capitalize on the enormous science potential with GWs from NS binaries systems will require accurately measuring both the GWs from the inspiral that determine the progenitor properties (e.g. masses, spins, cold NS matter, orbital eccentricity) and the GW signatures of the new physics encountered at the merger and its aftermath, as 3G detectors such as ET will enable. Figure ~\ref{fig:merger_phases} illustrates the potential of ET for detecting these effects, compared to current 2G detectors, for a NS binary at a typical distance of 100~Mpc (see also \cite{Andersson:2017kru}). The 2G detectors such as Advanced LIGO/Virgo are largely limited to observing the long inspiral, dominated by the center-of-mass dynamics of the NSs, with glimpses of the tidal effects which become important a higher frequencies, and are insensitive to the details of the merger and post-merger epochs. By contrast, a detector such as ET, besides observing the inspiral phase and the onset of tidal effects with much higher signal-to-noise ratio, will also clearly observe the final merger and post-merger signals and enable detailed insights into the fundamental properties of nuclear matter in a large swath of unexplored regimes in the QCD phase diagram. 

The coalescence events of NS-NS and NS-BH systems also have key significance as the production site of elements heavier than iron in the cosmos. Heavy elements can be synthesized from the neutron-rich material expelled during the merger or tidal disruption of NSs or through winds from the remnant accretion disk. The subsequent radioactive decay of the freshly synthesized elements powers leads to an electromagnetic transient known as a kilonova. Multi-messenger observations of a large sample of NS binaries will provide the unique opportunity to study heavy element formation at its production site, to determine how the initial conditions of an astrophysical binary system map to the final nucleosynthetic yields, and the extent to which different NS binary progenitors contribute to the cosmic abundances over time.

{\em In summary, the greater sensitivity and larger frequency bandwidth of a detector such as ET will be critical to observe a diverse population of NS binaries, accurately measure GW signatures of matter during the inspiral, and probe details of the merger and post-merger phenomena. These measurements are essential to substantially advance frontiers of subatomic physics by determining (1) the properties of cold, dense matter in NSs, (2) the new physics encountered during a binary NS merger, at higher temperatures and more extreme densities, and (3) the formation of heavy elements in the cosmos from synergies with electromagnetic observations.}

\subsubsection{Continuous waves from spinning neutron stars}

A spinning NS, isolated or in a binary system, can also emit continuous semi-periodic GWs if asymmetric with respect to its rotational axis~\cite{Maggiore:2018zz}. 
Such asymmetry can derive from frozen deformations produced right after its violent birth, from a strong enough inner magnetic field (provided it is not aligned with the rotation axis), from non-axisymmetric motions or density perturbations due, for instance, to Ekman flow or to the excitation of normal modes associated to the star rotation, such as the so-called r-modes, or due to thermal or composition gradients induced by matter accretion from a companion star~\cite{Lasky:2015uia}. 

No continuous gravitational wave signal has so far been observed  by Advanced LIGO and Advanced Virgo, see e.g.~\cite{Authors:2019ztc,Abbott:2019bed,Pisarski:2019vxw,Abbott:2018qee,2019arXiv191005097P,Abbott:2019uwg} for recently obtained upper limits.
{\em The detection of continuous GWs from NS by ET would be a fundamental breakthrough, that would provide  clues about  the condition of formation of isolated NS,  their spin, thermal evolution and magnetic field. Furthermore,  detecting such a signal would again give information on the inner structure of NS and on the corresponding aspects of nuclear and particle physics, such as the existence of exotic matter in the NS core.} In particular,
the maximum degree of deformation that a NS can sustain depends on the equation of state: for standard equations of state the maximum value of the ellipticity is $\epsilon_{max} \sim 10^{-6}$~\cite{Horowitz:2009ya}, but for exotic objects, containing hyperons or quark matter, is expected to be much higher, $\epsilon_{max}\sim 10^{-4}-10^{-3}$~\cite{JohnsonMcDaniel:2012wg}. In practice, it is difficult to predict the actual deformation of  a specific NS, that can depend on the star's history and could be well below the maximum sustainable value. A recent argument~\cite{Woan:2018xxx} suggests that the typical spin-down of millisecond pulsars can be explained assuming a typically ellipticity of about $10^{-9}$. 

\begin{figure}[t]
\centering
\includegraphics[width=0.75\textwidth]{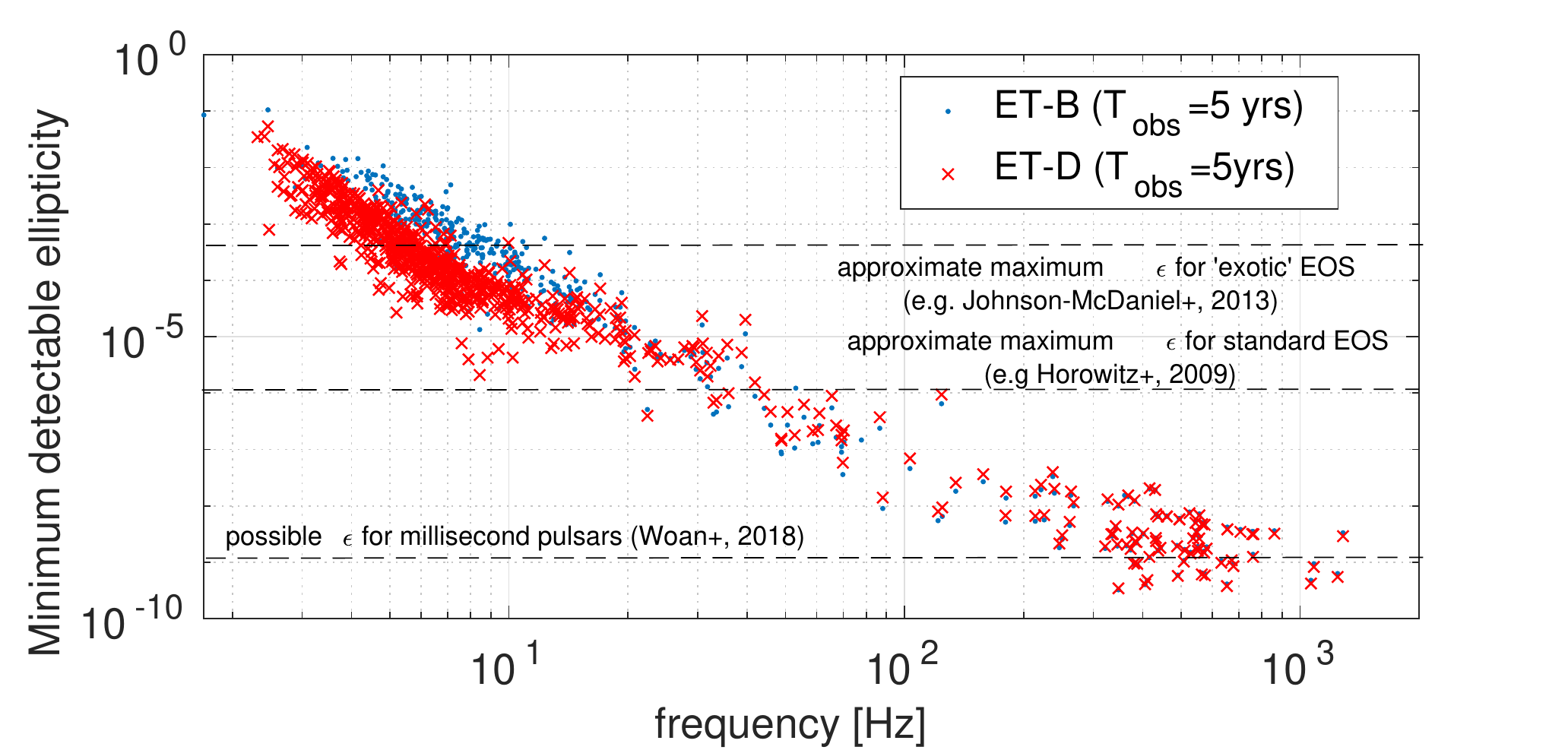}
\caption{Minimum ellipticity detectable by ET at 90$\%$ confidence level in a full coherent search of continuous waves from known pulsars, assuming an observation time $T_{\rm obs}=5$~yr. Two detector configurations, ET-B and ET-D, are taken into account. 
}
\label{fig:eps_vs_dist}
\end{figure}

Figure \ref{fig:eps_vs_dist} (which is an updated version of that shown in~\cite{Abernathy:2011}) shows the minimum detectable ellipticity (at 90$\%$ confidence level) for currently known NSs potentially emitting in the detector band, assuming two proposed ET configurations and that a full coherent matched filter analysis is done over an observation time $T_{\rm obs}=5$~yr; we see that ET will be sensitive to ellipticities of the order of few times  $10^{-10}$ for the nearest millisecond pulsars, and of $\sim 10^{-6}-10^{-7}$ for young pulsars. To produce this plot we used the minimum detectable signal strain amplitude at 90$\%$ confidence level, $h_\text{min}$, which is given -when matched filter is used- by $h_\text{min}(f)\approx 10\sqrt{S_\text{n}(f)/T_{\rm obs}}$, where $S_\text{n}(f)$ is the detector noise spectral density (measured in 1/Hz) at the frequency $f$. Hence, by exploiting the relation between the signal amplitude and the source ellipticity, at known distance and emission frequency (see eq.~5 of \cite{Abadie:2011md}), the minimum detectable ellipticity can be computed.   
{\em It is quite impressive to realize that detecting GWs due to an eccentricity of, say, $\epsilon=10^{-7}$ in a NS means that we would detect the effect  due to a ``mountain'' on a NS, with a  height of about $10^{-7}\times 10\, {\rm km}=1\, {\rm mm}$}
(or $10^{-2}$~mm, for $\epsilon=10^{-9}$).

\begin{figure}[t]
\centering
\includegraphics[width=0.75\textwidth]{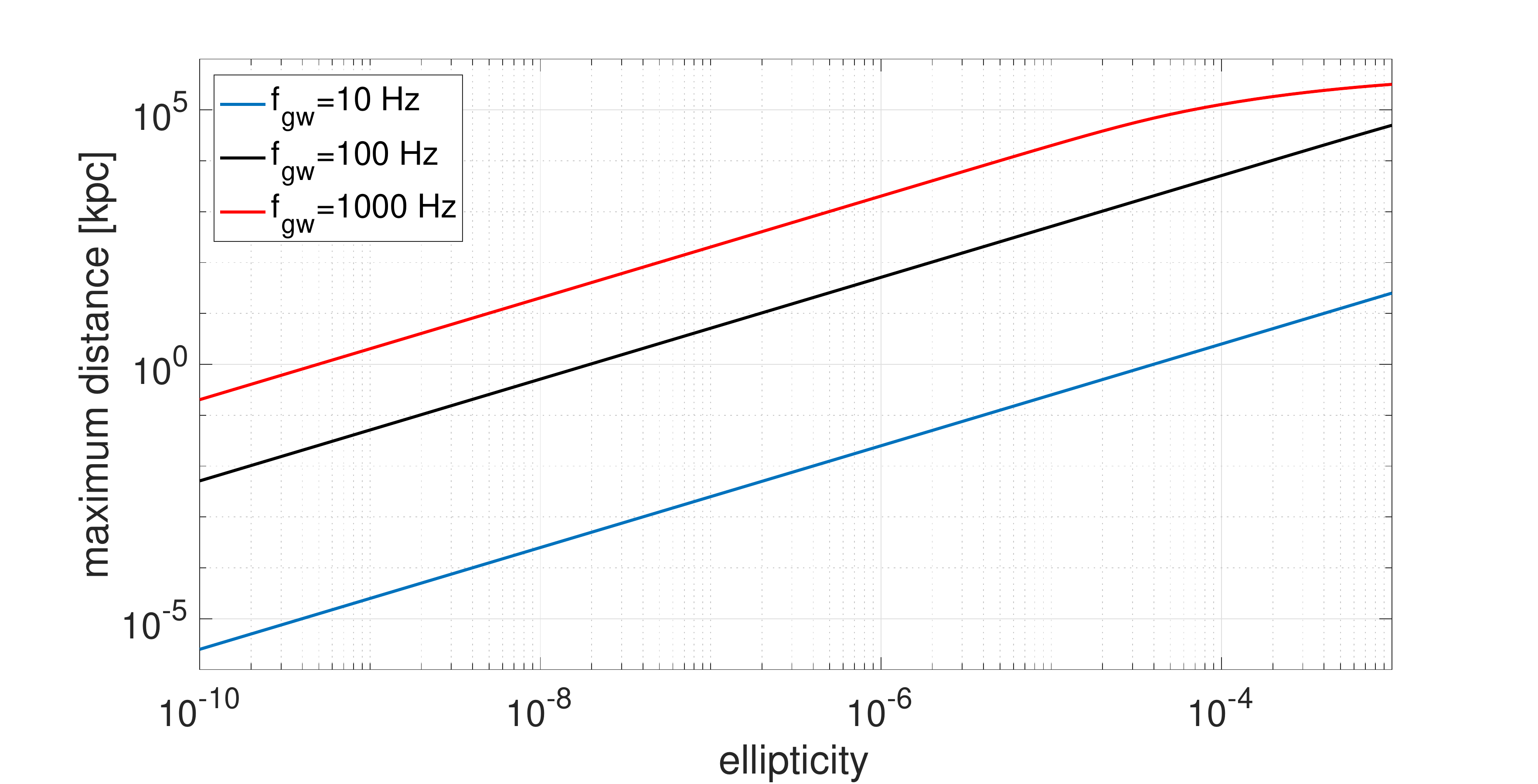}
\caption{Maximum distance at which a continuous wave source would be detected in a full coherent search over $T_{\rm obs}=5$~yr as a function of the source ellipticity, for three different values of the initial signal frequency, for the ET-B detector configuration. Here we make the assumption that source spin-down is dominated by the emission of GWs and that matched filtering is used in the analysis.
}
\label{fig:et_eps}
\end{figure}

Figure \ref{fig:et_eps} shows the maximum distance at which a continuous wave source would be detected, by making a full coherent search over an observation time  $T_{\rm obs}=5$~yr, as a function of its ellipticity for different values of the signal frequency, assuming that the source spin-down is dominated by the emission of GWs. In this case by equating the signal amplitude to the minimum detectable value $h_\text{min}$, we can obtain the maximum distance as a function of the source ellipticity for any given initial value of the signal frequency. It is worth to stress that a neutron star spinning at say 50 Hz (and then emitting a continuous GW signal at 100 Hz) would be detectable in the whole Galaxy as long as its ellipticity is larger than $10^{-7}$. 
Very  fast spinning  and highly distorted neutron stars,  such as newborn magnetars produced in core collapses or as post-merger remnants of coalescing binaries, could lead to detectable emission at even higher frequencies. In this case the signal can only be observed for a shorter time, since these objects  
are characterized by a very high spin-down and the signal frequency eventually leaves  the detector sensitivity band within a  few days. However, at birth they could have ellipticities as large as $10^{-3}$~\cite{DallOsso:2018dos} so,  even taking into account uncertainties in the data analysis due to the very large initial parameter space (initial frequency, spin-down, braking index), these objects could still  could still be detected out to distances of tens of Mpc~\cite{Miller:2018rbg}. 

The detection of continuous signals from spinning neutron stars will represent a complementary tool to the merger and post-merger signal for the study of the NS interior, especially if concurrent electromagnetic observations are available. It will also allow provide clues about NS formation and demography, their spin, thermal and possibly magnetic field evolution.  

It is also important to observe that, even for a single detector such as ET, the prolonged observation of the continuous signals from a NS allows, over the time of months, to localize it with extremely good accuracy, exploiting the movement of the Earth during the observation time. The angular resolution, for a source emitting a signal at frequency $f_\text{gw}$, and searched over a timescale $T_\text{obs}$, is about $3\times 10^{-6}\left(100~\text{Hz}/f_\text{gw}\right)\left(1~\text{yr}/T_\text{obs}\right)$ rad~\cite{Astone:2014esa}.

Low-Mass X-ray Binary (LMXB) are another very interesting target for ET, under the assumption that a balance among angular momentum accreted through matter infall and the emission of GWs exists. In this scenario, by comparing the signal amplitude given in \cite{Bildsten:1998ey} to the planned ET sensitivity curve, we see   that at least one LMXB, namely Sco-X1, would be detectable by ET if the unknown star rotation frequency is in the range from a few Hz to about 1 kHz.

\subsubsection{Burst signals from neutron stars}
Neutron stars can also emit transient bursts of GWs in association, for instance, to magnetar giant flares and pulsar glitches. Magnetars are NSs endowed with a very strong magnetic field of $10^{14}$ G or more, and are observed as anomalous X-ray pulsars (AXP) or soft gamma-ray repeaters (SGR)~\cite{Duncan:1992hi}. SGRs are characterized by recurrent short-duration X-ray bursts and more energetic giant flares ($10^{44}-10^{47} {\rm erg}\,  {\rm s}^{-1}$ in $\sim 0.1$~s), due to global rearrangement of the inner magnetic field or of the magnetosphere. These events can induce a significant structural changes in the NS,  or excite polar oscillations, like the f-modes, causing an emission of GWs. The
strongest pulsar glitches, like those of the Vela pulsar, are explained as due to the occasional ``unpinning'' of quantized superfluid vortices in the interior of the spinning-down NS, which move outward and release their angular momentum. The prediction on the emission of GWs are, however, still not robust due to the lack of a detailed knowledge of the process. Moderately optimistic models predict that these signals can be detected by ET.

\subsection{Multi-messenger astrophysics: synergies with other GW detectors and electromagnetic/neutrino observatories}
\label{sec:MM}

ET, with its triangular configuration corresponding to three nested interferometers, is designed so to have an extraordinary science output even when operated as a single GW detector. However, a further enhancement of its capabilities will take place when making use of  the synergies with other detectors that could be operating at the same time.

\subsubsection{Networks of  gravitational-wave detectors}\label{sect:net3G}

The first obvious synergy is with other GW detectors, either with a network of second-generation (2G) GW detectors (Advanced LIGO Livingston, Hanford and India, Advanced Virgo, and KAGRA),  
or with another
third-generation detector such as  Cosmic Explorer (CE), currently under study in the US~\cite{Dwyer:2014fpa,Reitze:2019iox} (possibly in a network configuration with one ET and two CE detectors).   A first important benefit from coincident detections in different detectors
will be the reduced impact of instrumental glitches on GW detections and analyses.\footnote{Actually, a single ET triangle might already provide a good enough background rejection through its null stream. However,  this is not 100\% efficient due to possible correlations between instrument noise in different ET interferometers.}
These problems affect, in particular, short duration transients lasting a few milliseconds - especially the unmodeled ones - while it will be less problematic for the signals associated to the coalescence of BHs or NSs binary systems, thanks to the relatively long duration of the signals in the sensitivity band of the detectors.

A network of detectors, compared to a single detector, will also improve significantly the accuracy in the localization of the sources. As we already mentioned, for the continuous GWs emitted by spinning NSs a single ET detector  already provides a very accurate parameter estimation - including position - thanks to the very specific modulation of the signals due to Doppler effect induced by the Earth motion.   For coalescing NS-NS binaries a single ET detector  still has some localization capability since, for a low-mass system such as a binary NS, the signal can stay in the detector bandwidth for a long time,  of order of a few days, and again the modulation due to  the Earth motion allows us to localize the source. In this case  an average angular resolution would be around  $200\deg^2$\, for a binary NS at $z=0.1$, but can become of order of just a few $\deg^2$ for the best localized 
sources~\cite{Zhao:2017cbb,Chan2018}. No significant localization will in general be available for the  vast majority of BH-BH and BH-NS binaries, that, because of their higher total mass, will stay in the detector bandwidth for a much shorter time.
A  network of three 3G GW detectors, in contrast, will have quite  good localization accuracy  for all types of binary coalescences; for example, a large fraction of NS-NS binaries will have sky localization smaller than 
1 deg$^2$ up to $z=0.5$~\cite{Hall:2019xmm}. 
In terms of science output, this means that a 3G detector network will be able to provide good localization information to electromagnetic observatories, making possible the detection of an
electromagnetic counterpart. However, as we will discuss in more detail in Section~\ref{sec:MM-EM},
even for the largest telescopes operating in the optical or infrared, such as the 39-m E-ELT~\citep{ELTinstruments}, it will be difficult to observe the kilonova associated to a NS-NS binary beyond $z\sim 0.5$ photometrically, and beyond $z\sim 0.3$ spectroscopically \citep{Chornock:2019rrt,Clenet2013} (see for example \url{http://www.mosaic-elt.eu/index.php}). 
Furthermore, the direct detection with these large telescopes will be possible only for arcmin-arcsec localized sources.
For more distant NS-NS mergers,  the electromagnetic counterpart detection would be possible through the associated emission of gamma-ray bursts. Accurate localization would
be also important in order to use statistical techniques, such as those based on the probabilistic determination of the redshift of the associated galaxy~\cite{Schutz:1986gp,DelPozzo:2011yh},  as well as on cross-correlations between binary coalescences and large-scale structures~\cite{Mukherjee:2018ebj,Scelfo:2018sny}.
These statistical techniques will be important for the cosmological applications of coalescing NS-NS binaries that will be discussed in Section~\ref{sec:cosmos}. 

On the other hand, the different sensitivity curves planned for  ET and CE imply that, from other points of view, these detectors will be complementary. For instance,  as can be seen from  Fig.~\ref{fig:gw_horizons}, ET will be  able to detect heavier systems, with total masses higher than $10^3~\Msun$ (thanks to its sensitivity in the low-frequency regime), while CE has a greater reach for light systems such as NS-NS binaries. The different sensitivity curves also mean that, for a given astrophysical system, the signal-to-noise ratio is accumulated differently in ET and in CE, providing complementary information.

\begin{figure}[t]
\centering
\includegraphics[width=0.48\textwidth]{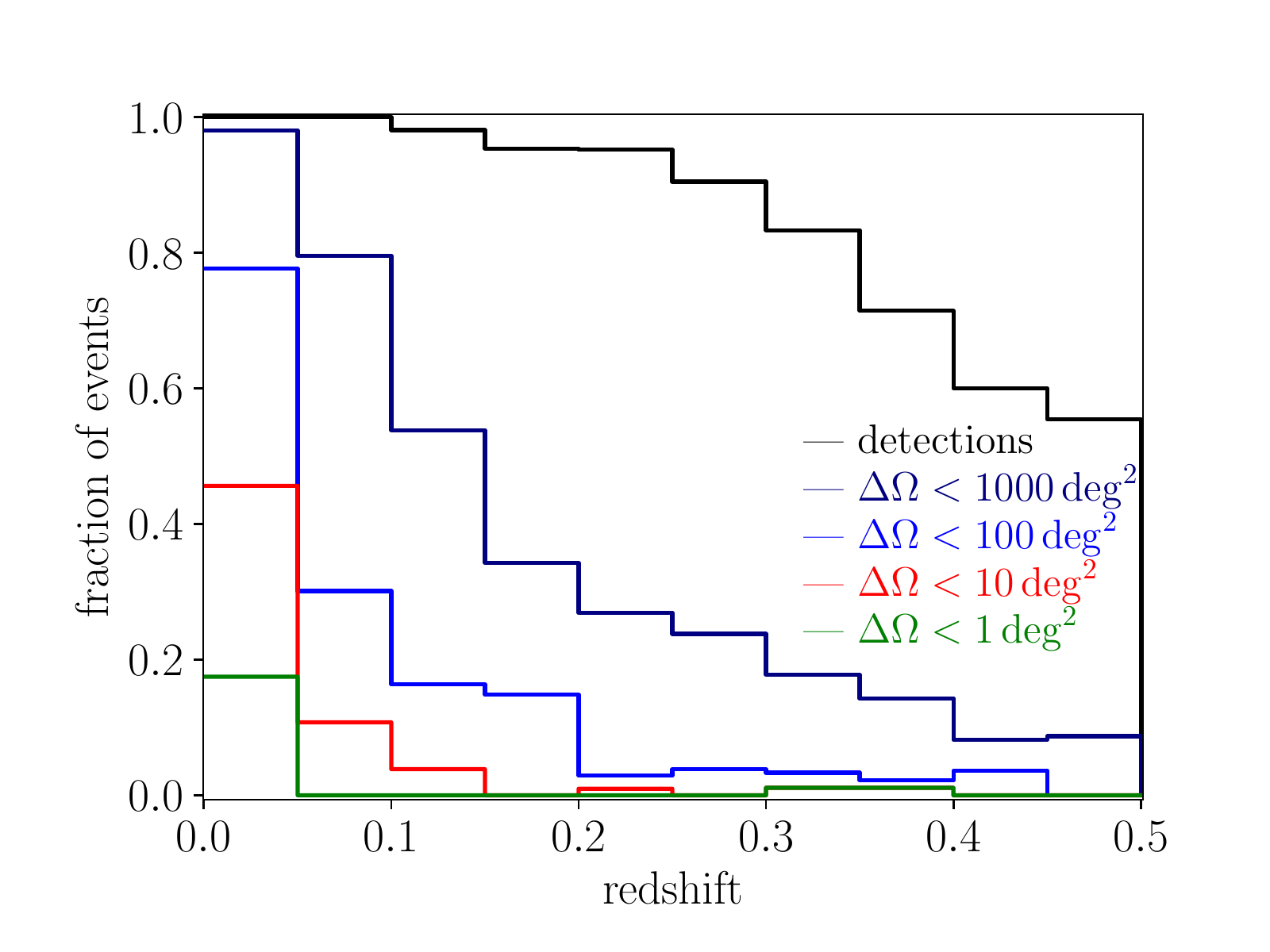}
\includegraphics[width=0.48\textwidth]{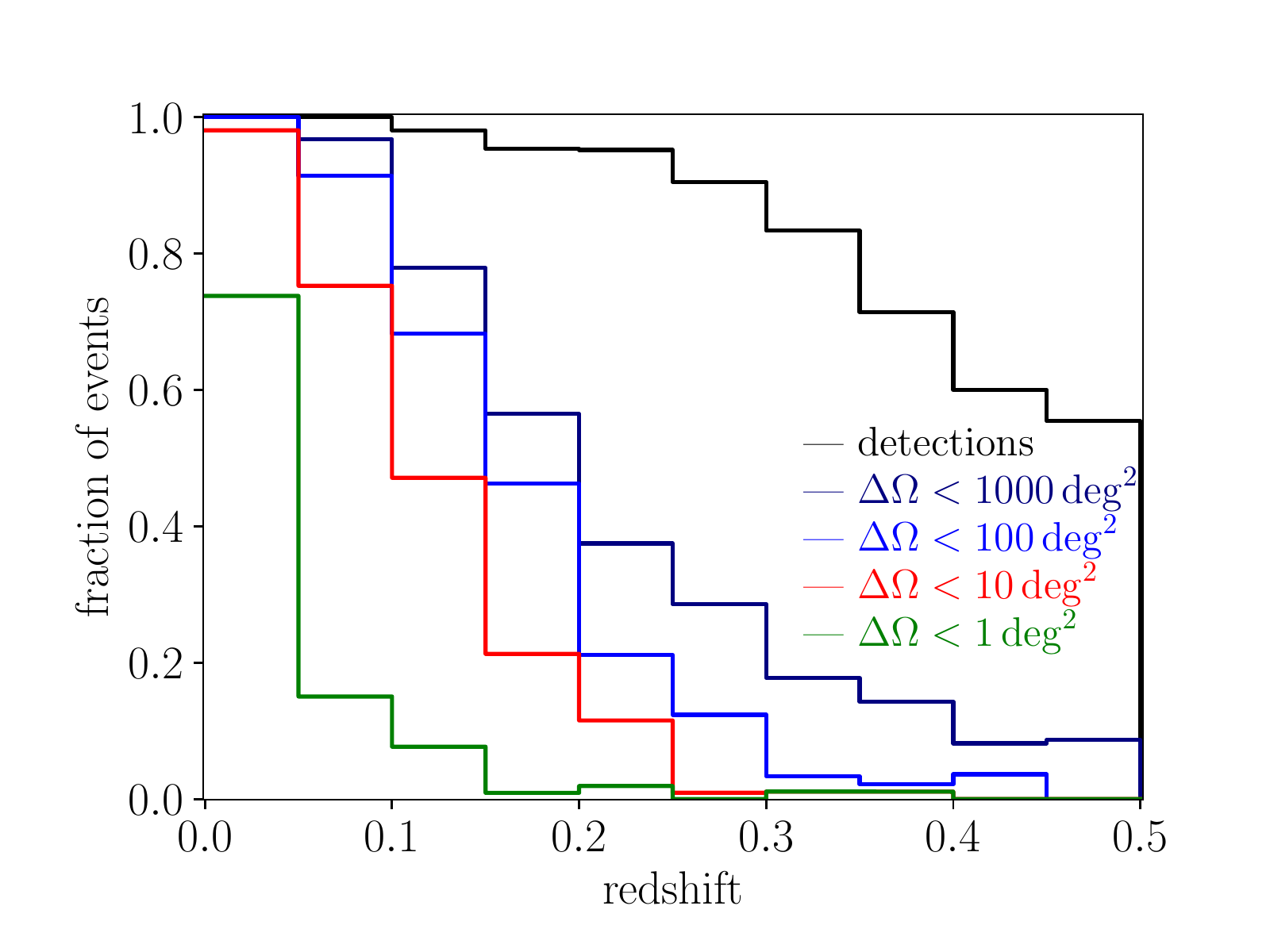}
\includegraphics[width=0.48\textwidth]{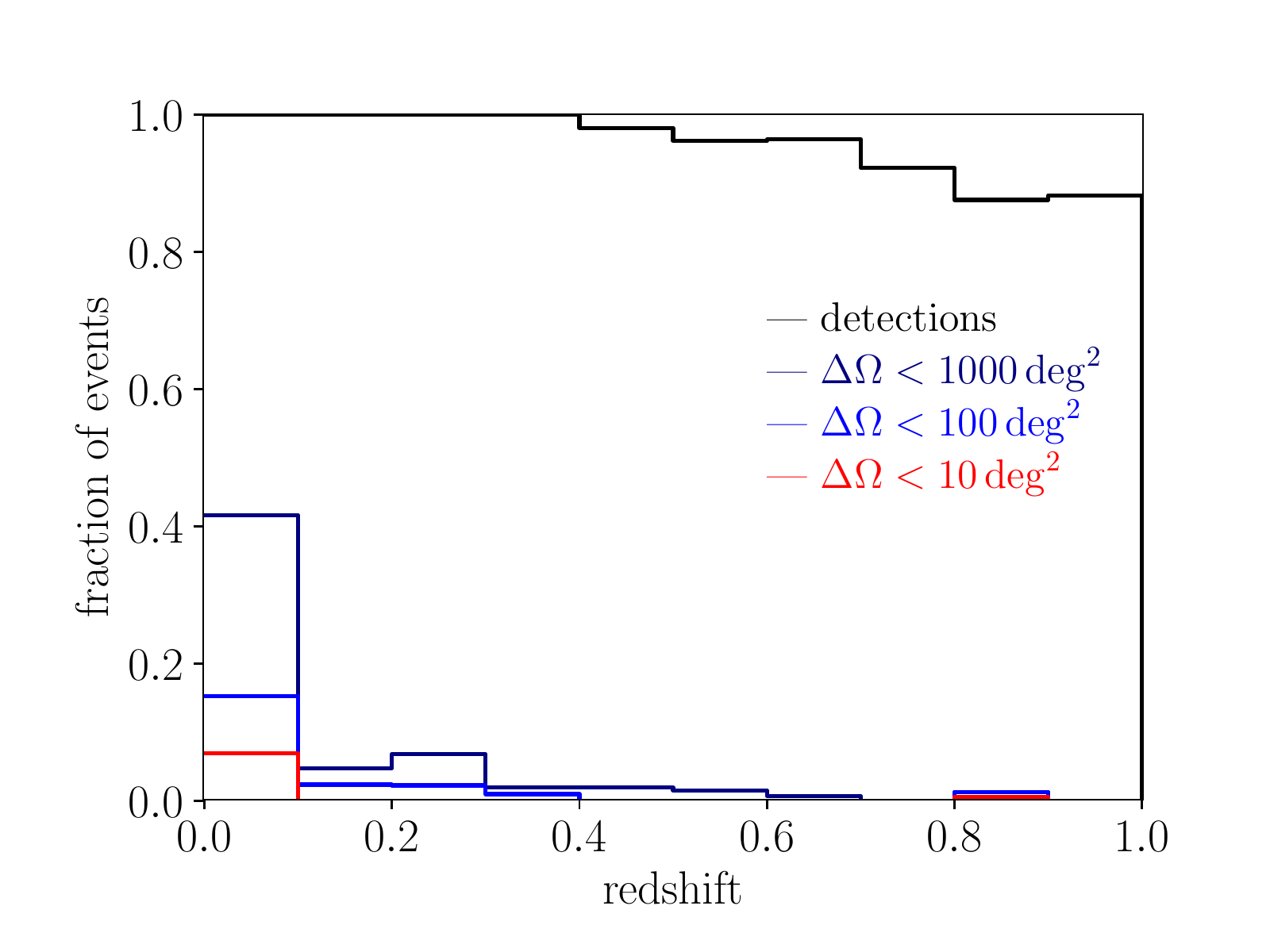}
\includegraphics[width=0.48\textwidth]{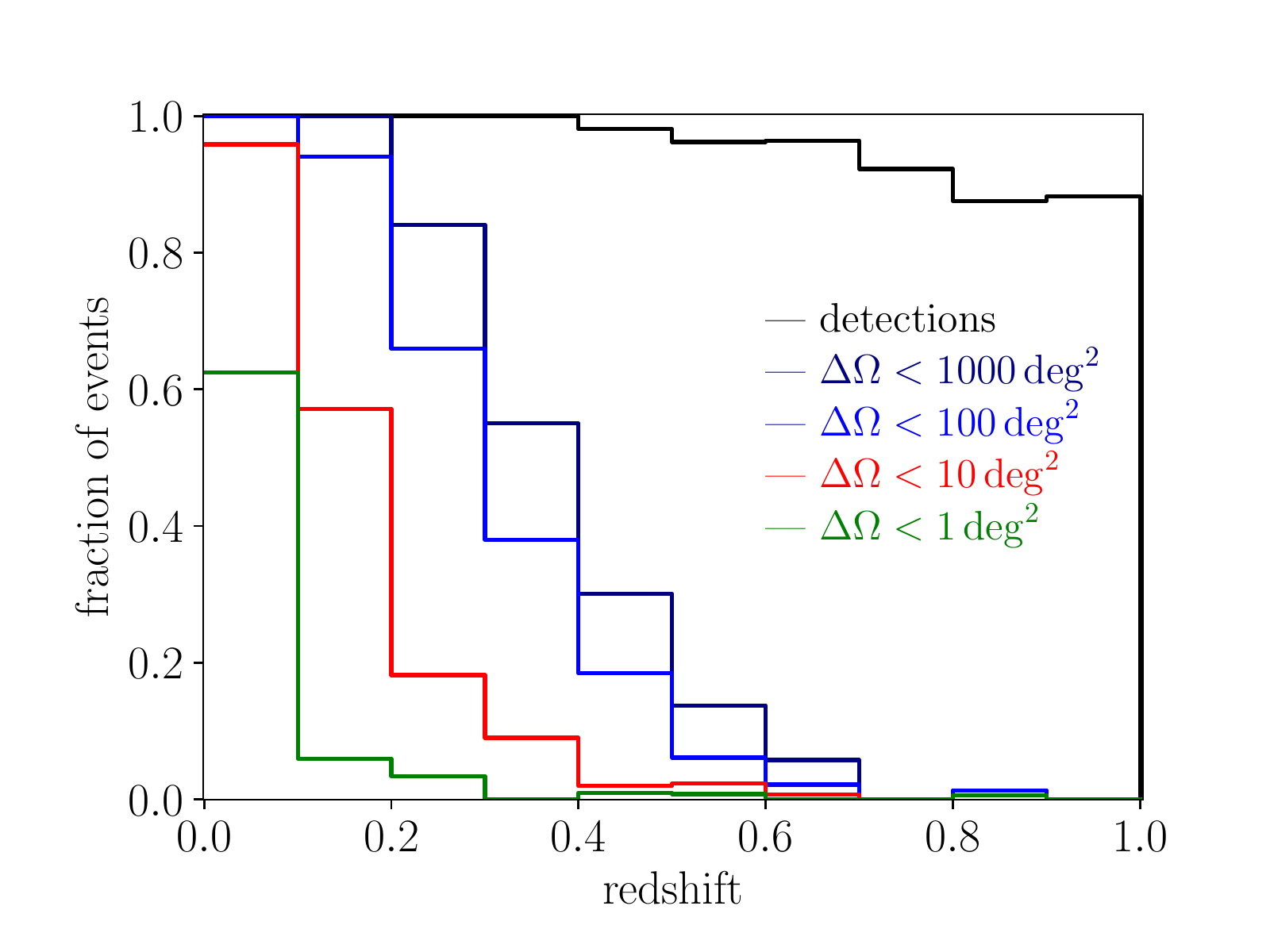}
\caption{Localization errors of BNSs and BBHs per redshift bin. Top: BNS. Bottom: BBH. Left: ET alone. Right: ET together with currently foreseen future upgrades of Advanced LIGO Hanford, Livingston, India, Advanced Virgo, and KAGRA. We impose a conservative ET detection threshold SNR of 12, and for the plots in the right column that at least one 2G detector has SNR $> 4$ to include 2G data for the sky localization. Duty cycle is taken 100\% for all detectors.}
\label{fig:gw_ET2G}
\end{figure} 

ET is also expected to operate with a network of five second-generation (2G) GW detectors: Advanced LIGO Livingston, Hanford and India, Advanced Virgo, and KAGRA. Assuming currently foreseen sensitivity increase of 2G detectors from the implementation of technology upgrades~\citep{Aasi:2013wya}, the improvement of the sky localization of this network compared to ET alone is shown in Fig.~\ref{fig:gw_ET2G}. The top row is for binary NSs, while the bottom row is for binary BHs. We obtained these results from a Fisher-matrix analysis based on a time-domain simulation of the GW signals to incorporate the effect of a rotating Earth, which is fundamental to the sky-localization capability of a single ET detector. Accordingly, about 80\% of BNS mergers in $0.05 < z < 0.1$ will be localized within 10\,deg$^2$ in such a network, while ET alone achieves for a similar fraction of BNS mergers better than 1000\,deg$^2$ localization errors in the same redshift bin. Note that the 2G network has a significant impact on sky-localization of BNSs up to redshifts of 0.3. It is also remarkable that a single ET detector is able to provide good sky localization at least for a small fraction of the closest BBHs.

\subsubsection{Joint gravitational and electromagnetic observations}
\label{sec:MM-EM}

The discovery and electromagnetic follow-up of GW170817 showed the enormous potential  of gravitational-wave observations for multi-messenger astrophysics. The gravitational-wave observations combined with the results from the extensive multi-wavelength observational campaign (still undergoing) had a huge impact on our knowledge of the physics of compact objects, relativistic jets, nucleosynthesis, and cosmology \citep[see e.g.,][]{Abbott:2019,Radice:2019,Mooley:2018, Ghirlanda:2019,Cote:2018,Kasen:2017,Abbott:2017xzu,Hotokezaka:2019}. Identifying the electromagnetic signatures of the gravitational wave sources enables to maximize the science return from a gravitational-wave detection by probing  strong-field gravity,  matter under extreme conditions together with the interaction of the source with its environment.  

ET will detect binary systems containing neutron stars up to redshifts corresponding to the peak of the cosmic star-formation rate. This represents a step forward for studying the formation, evolution and physics of neutron stars in connection with kilonovae and short gamma-ray bursts, along with the star formation history and the chemical evolution of the Universe. Its better  sensitivity and larger sample of detections will make possible to connect progenitor and merger remnant properties to the multi-messenger emission mechanism, and 
to distinguish between geometric and intrinsic properties of the source. ET will operate with a new innovative generation of observatories covering from the radio to the high-energy bands. In the following we will not discuss exhaustively all the instruments that could operate with ET, but we will consider some of the main future observatories with a large involvement of the European community.

Sensitivity (in terms of observable distances) and sky-localization capabilties of ET will determine the observatories that can effectively
operate in synergy with ET. Larger sky localization and larger distance decrease the efficiency of the electromagnetic search, due to a major difficulty to find a fainter counterpart among many contaminant transient signals and many possible host galaxies. ET as single detector can localize the majority of detectable binary neutron stars at a distance of $\leq 200$ Mpc to within a $90\%$ credible region of 100 ${\rm deg}^2$~\cite{Zhao:2017cbb,Chan2018}.  Operating with a network of five 2G detectors, similar localization capabilities will be possible up to about 1 Gpc. Going to larger distances,  the sky-localization of ET will be larger than $10^3-10^4 \,{\rm deg}^2$ for the majority of the sources. These localizations require instruments able to monitor quickly large parts of the sky,  and follow-up observations to characterize the nature of the counterpart. The Square Kilometre Array \cite[SKA,][]{Carilli2004}, the Large Synoptic Survey Telescope~\citep[LSST,][]{Ivezic2008}, THESEUS~\citep[mission concept,][]{Amati:2017npy}, the Cherenkov Telescope Array~\citep[CTA,][]{Acharya2013} will be able to observe large regions of the sky from the radio, optical to the X-ray and very high energy, going to deeper sensitivity than current observatories; 40-meters class telescope, such as the European Extremely Large Telescope~\citep[E-ELT][]{Clenet2013} and satellite like the Advanced Telescope for High ENergy Astrophysics~\citep[ATHENA,][]{Nandra2013} will be able to characterize the source in the optical and X-ray band. 
For specific emissions and science goals, there will be two regimes: close sources (up to 400 Mpc), where we will strongly benefit of the higher sensitivity of ET and thus, with respect to 2G detectors, we will be  able to better evaluate the properties of the source progenitor, merger remnant and environment interactions in single detections;  and a second regime for distant sources where we will benefit of larger samples of joint electromagnetic and gravitational wave detections.

The ET sensitivity at low frequencies enables enough signal-to-noise ratio to accumulate before the merger, making possible an early detection and warning for the electromagnetic/neutrino followup. Requiring a signal-to-noise ratio  $\geq 12$ and a sky localization smaller than 100 deg$^2$, ET can send an early warning alert between 1 and 20 hours before the merger (with the mean of the distribution at about 5 hours) for signals at 40 Mpc \cite{Chan2018}. At 200 Mpc, about 30\% of the detectable signals had accumulated enough SNR for early warning between 1 to 6 hours prior to the merger. About 10\% of the detectable sources within 400 Mpc can still be announced with an early warning smaller than 1 hour; this percentage drops to 3\% for sources at 800 Mpc, and  no detection of source at 1.6 Gpc can be announced before 1 hour from the merger. Releasing an early warning increases the chance of successfully detecting the electromagnetic counterparts, by allowing astronomers to point the telescope in the region of the signal or starting the monitoring to obtain pre-merger images, which are extremely useful to discard contaminant transient objects. This enables to detect the early electromagnetic emission, which is fundamental to understand the physics of the engine and the merger remnant.


\subsubsection{Relativistic astrophysics and short gamma-ray bursts}

The cosmic origin of elements heavier than iron has long been a mystery. 
GW17017 provided the first observational test to theoretical models which predict the rapid neutron capture process (r-process) nucleosynthesis during  binary neutron star mergers \citep[see e.g.,][]{Metzger2017,Kasen:2017,Tanaka2016}. 
The thermal emission observed in the ultraviolet, optical, and near-infrared detected with GW17017 was found to be consistent with kilonova emission powered by the radioactive decay of heavy nuclei (including lanthanides) synthesized in the merger ejected by the r-process \citep[see e.g.,][]{Pian:2017,Smartt:2017,Evans:2017}.  
On the basis of the merger rate estimated using the LIGO and Virgo observations and the amount of ejected mass estimated by the kilonova observations, binary neutron star mergers are now understood to be a major channel of r-process production, able to explain the heavy elements abundances in the Milky Way stellar population. However, it is still uncertain the role of rare classes of supernovae, such as the collapsars associated with long gamma ray bursts, which are expected to be an additional significant source of r-process elements \cite{Siegel2019}.
It is also still incomplete the interpretation of the kilonova emission and spectral evolution over many days; the contribution of the multiple ejecta (dynamical, wind, viscous, etc.) and their properties are not fully understood, as well as what are the exact elements formed and their abundance, and what is the role of the merger remnant. Only a larger sample of kilonova, possibly extending to larger distance, will enable us to probe the details of the kilonova emission mechanism, and the role of binary neutron star mergers in the Universe enrichment of heavy elements along the cosmic history.

When ET is expected to observe the sky, LSST will operate as a wide field-of-view survey able to detect kilonova emission up to 800 Mpc. 
Up to the same distance, photometric and spectroscopic characterization will be possible using ground-based 30--40 m telescopes such as the Thirty Meter Telescope \citep[TMT,][]{Skidmore2015} and E-ELT, and the James Webb Space Telescope \citep[JWST,][]{Gardner2006}.  The binary neutron star mergers detectable in this volume are of order $10^3$ per year. Among them, a few hundred are expected to be in the LSST surveyed field. For the majority of these sources, the gravitational-wave localization uncertainty by ET will make it difficult to identify the optical counterpart among many optical transient contaminants. However, a significant number of joint GW/kilonova detections (of order of several tens) becomes possible considering ET operating with the second generation of detectors. For joint gravitational wave/kilonova detection, the precision of parameter estimation for the progenitor system (total mass, mass ratio, spin, and neutron star tidal deformability) and the detection of the signal from the merger remnant made possible by ET represent an unprecedented opportunity to understand the physics governing the kilonova emission, and the nature and equation of state of neutron stars.

A single ET detector, even in the absence of good source localization, will still be able to perform joint observations with gamma-ray burst (GRB) detectors, through  the observation of a temporally coincident GRB. In turn, this  can  allow for the measurement of the redshift of the source when the high-energy satellite is capable to precisely  localize the source. Indeed, GRB satellites such as \emph{Swift} regularly alert ground based spectrographs to obtain the redshifts of the host galaxies of the detected GRBs. The study in
\cite{Stratta:2017bwq,Belgacem:2019tbw} indicates that several tens  short GRB  per year will be detected by a GRB mission such as THESEUS in coincidence with a gravitational signal in ET. Beside the collimated prompt GRB emission, more isotropic soft X emission is also expected from the afterglow. This could lead  possibly to a few hundred more coincident detections per year~\cite{Stratta:2017bwq}. 

The discovery of the gamma-ray emission  associated with GW170817 and the following afterglow observations significantly improved our knowledge of short GRB jets.
Approximately two seconds after GW170817, the Fermi space telescope detected a weak short-duration gamma-ray burst, GRB170817A. Even if it showed the classical observational features that led to classify it as a short GRB, its total gamma-ray energy of about $10^{46}$ erg was many orders of magnitude smaller than the typical energy of any GRB observed before~\citep{Goldstein:2017mmi,Savchenko:2017ffs}. Nine and sixteen days after the  GW observation of the merger, X-ray and radio emissions were also detected~\citep{Troja:2017,Hallinan:2017}. Over longer timescale the radio, optical, and X-ray observations showed a slow achromatic flux increase until about 150 days before starting to decline~\citep{Lamb:2018qfn,Dobie2018,Davanzo2018,Margutti2018}. High-resolution radio observations~\citep{Ghirlanda:2019,Mooley:2018} were able to constrain the source size and to show a source displacement consistent with the launch of a jet which successfully breaks through the ejecta developing an angular structure, i.e. a narrow ultra-relativistic jet surrounded by less-collimated and slower material. The  structured jet was observed off-axis (i.e. the observer was misaligned with respect to the collimated ultra-relativistic jet). However, while multi-wavelength observations over two years have built a broad consensus about the interpretation of the non-thermal afterglow emission, the origin of the extremely faint prompt gamma-ray emission observed far from the jet core is still under debate; a gamma-ray emission arising from the slower part of the jet or a gamma-ray emission due to a cocoon shock breakout~\citep[see e.g.,][]{Beniami2019,Salafia2019,Kathirgamaraju2018,Nakar2018,Kasliwal2017}.

These  results have been extraordinary. Nevertheless, only a detector such as ET will  have the unprecedented capability to completely probe short GRB jet properties, by exploring up to high redshift a large population of neutron star mergers observed perpendicular to the orbital plane (on-axis) and off-axis. Mission concept such as THESEUS will enable to detect $20-40$ on-axis short GRB/year with a localization accuracy of 1-5 arcmin up to a redshift of 5~\cite{Amati:2017npy,Stratta:2017bwq}.  After each detection, the rapid alert system will enable to point ground-based spectrographs, such as the ones in E-ELT, and satellites such as ATHENA. THESEUS will give the precise position of the source, and ET and the multi-wavelength follow-up will allow us to connect detailed information of the progenitors and merger remnant properties to the jet and environment properties. It will be possible to build a statistical sample of binary neutron star mergers  able to probe the shape of the jet structure, if it is universal, and what is the typical opening angle for short GRBs. It will be possible  to constrain the  luminosity function  of short GRBs and its relation to the jet structure and the intrinsic luminosity evolution, and to understand what is the efficiency of the jet to break through the material surrounding the NS-NS mergers. ET will be crucial to identify the nature of the binary neutron star merger remnant (black-hole, unstable or stable neutron star) and how this is connected to the short GRB central engine and afterglow properties. Finally, large sample of GW/GRB will clarify the role of NS-BH binaries as progenitor of short GRBs. 

ET will guarantee that instruments such as THESEUS will have a gravitational-wave detection for each detected on-axis GRB. Over a few years, it will be possible to build a sample of tens to $O(100)$  joint detections with luminosity distance measured by gravitational-waves and  redshift measured by ground-based telescopes, such as VLT and ELT. These detections will provide precise measurements of the Hubble constant, helping to break the degeneracies in determining other cosmological parameters obtained by CMB, SNIa and BAO surveys, and to study the nature of dark energy~\cite{Belgacem:2019tbw};  see section~\ref{sec:cosmos} for details.

The detection of a faint off-axis gamma-ray signal such as the one observed by Fermi and INTEGRAL for GW170817, will be difficult for present and the planned future detectors at distances larger than 100 Mpc. However, a fraction of NS-NS merger are expected to produce long-lived neutron stars. In this case, soft X-ray transient can be powered by the new-born neutron star spin-down emission. Even if never observed so far, this emission is expected to be powerful and nearly isotropic \cite{Metzger2014,Siegel2016}. Large field of view instruments, such as the one on board of THESEUS, will allow us to detect the brighter emissions up to 1 Gpc, thereby increasing the numbers of joint GW/electromagnetic detections in high-energy to a few hundreds per year. 

\subsubsection{Multi-messenger observations and core collapse supernovae}

Despite the remarkable progress of the theory, the explosion mechanism of supernovae is still an open question, and being able to measure the dynamics of matter at the onset of the explosion would bring invaluable information to the understanding of the physics of gravitational core collapse. 
What is fairly known since the 1970's is the role of the neutrinos in the explosion mechanism~\cite{CoWh66,Wilson74}. During collapse, the stellar core becomes opaque to neutrinos, producing a degenerate sea of trapped neutrinos within it, which subsequently diffuses out of the core on a timescale of order tens of seconds as the nascent proto-neutron star  cools and deleptonizes.
The three-flavor neutrino flux emanating from the proto-neutron star could power itself a core collapse supernova via neutrino heating on delayed timescales of order one second~\cite{JaMu96,mezz:98,MuJaMa12,MuJaHe12,takiwaki:12,hanke_13,murphy:13,CoOt13,cerda_13,MuMeHe17,radice:17,OcCo18,Powell:2018isq}.
This phenomenon is central to most models today, with the exception of models of rare events involving significant rotation, which may be powered magneto-hydrodynamically and where the dynamics proceeds  on shorter timescales~\cite{ZwMu97,DiFoMu01,DiFoMu02,dimmelmeier:07,ott:07,dimmelmeier:08,kuroda:14,yakunin:15,kuroda:16,andresen:17}.
Besides, the Standing Accretion Shock Instability (SASI)~\cite{blondin:03,foglizzo:07} which is an instability of the supernova shock wave itself is expected, in conjunction with turbulence, to modulate the accretion flow, to excite the proto-neutro star high frequency oscillation modes as well as to generate low frequency (100 $-$ 250 Hz) GWs~\cite{MuRaBu04,kuroda:16,andresen:17}. This low frequency GW emission depends on the nuclear equation of state, with softer equations of state leading to larger SASI amplitude~\cite{kuroda:16}.   

Core collapse SNe are not a guaranteed source for ET. A general consensus from all modern numerical simulations is that the expected GW signal is weak (GW released energy of the order of $10^{-9}$~$M_\odot c^2$)~\cite{Radice:2018usf}. Furthermore, the likely diversity of the GW emission mechanisms that are at play in SN explosion makes quite difficult to use matched filtering techniques for digging the signal out of the noise, contrary to what can be done with coalescing binaries or spinning neutron stars. As a consequence,
the detection of a core collapse supernova GW signal is very challenging
and the discovery horizon of the current 2G detectors is limited to our galaxy~\cite{gossan:16}. The expected galactic rate of type II/Ib supernova is also rather small ($\sim 1$ per 30 years)~\cite{Diehl:2006cf}. ET will extend the reach to our galactic neighborhood, so that the expected rate is such that, while detection is not assured, still it is a realistic possibility.  
A detection of the GWs emitted in core collapse would be a milestone, revealing the inner mechanisms of core collapse and opening remarkable perspectives in multi-messenger astronomy. In particular,
the neutrino emission which will be in coincidence, within few milliseconds, with the GW emission, should be detected by the current and future low energy neutrinos detectors (Super-K/Hyper-K~\cite{Abe:2016waf}, DUNE~\cite{Ankowski:2016lab}, JUNO~\cite{Lu:2014zma}, IceCube~\cite{Abbasi:2011ss}, the LVD~\cite{Agafonova:2007hn}, Borexino~\cite{Cadonati:2000kq} and KamLAND~\cite{Tolich:2011zz}) with a higher signal to noise ratio than the GW signal and a very precise time resolution (few milliseconds) which is a fundamental information to search for a  signal with low signal-to-noise ratio  in the GW data. The false alarm rate of GW searches can be significantly improved with the temporal localization given by the neutrino signal. Furthermore, there exists a strong correlation between the GW and neutrino signals as they are produced at the same interior location and will be powered by the downward accretion plumes associated with hydrodynamic instabilities present in the post-shock flow. These plumes and instabilities will modulate both signals~\cite{Tamborra14,BMuller14,Kuroda:2017trn}.
If the signal is likely to remain short (of the order of 1~s), it is expected to be wide band (from few Hz up to several kHz), with very different mechanism in each frequency band. The low frequency and high frequency ET conception design is very well suited for detecting such kind of GW signal.

\subsubsection{Multi-band GW observations with LISA}

Another potential very interesting synergy could take place with the space interferometer LISA~\cite{Audley:2017drz}, should ET be operational  by the time that LISA will fly (scheduled for 2034, for a nominal four year duration, with possible extensions), or up to several years  after the end of the mission. For most sources, ground-based 3G detectors and LISA are highly complementary, because of the very different frequency range where the operate, so, both for astrophysics and for cosmology, the scientific targets of ET and of LISA are in  general quite independent. 

However, for stellar-mass BH-BH binaries, interesting synergies are possible.
In particular, from the rate of BH coalescences inferred by the Advanced LIGO/Virgo O1 and O2 runs, one estimates that  LISA could detect  $O(100)$ stellar-mass BH-BH  binaries during their inspiral phase \cite{Audley:2017drz}, up to $z\simeq 0.4$.   Several years later (typically, 5 to 10 yr),  some of these events will cross into the ET window, where they will coalesce. For instance, the first observed GW event, GW150914,  would have been  in the LISA bandwidth from  about 10 yr to less than 1 yr before coalescence, and 
could have accumulated in the LISA bandwidth a signal-to-noise ratio between three and fifteen (depending on the detector configuration), and therefore might have been detected, if  at that time LISA had already been  in orbit~\cite{Sesana:2016ljz}. This could allow for multi-band GW observations, i.e. the observation of GW signals in widely different frequency bands. 

The actual possibility of  multi-band detections depends sensitively on a number of issues, in particular on the sensitivity curve assumed for LISA, particularly at high frequency, on  the details of the data analysis in LISA, and on the  mass distribution of the component BHs in  a BH-BH binary. The recent analysis in \cite{Caputo:2020irr} finds that LISA, in a 4~yr mission, will detect several tens of  stellar-mass  BH-BH binaries, consistently with the earlier estimates in~\cite{Sesana:2016ljz,Tamanini:2019usx}.\footnote{Smaller estimates have been obtained in~\cite{Moore:2019pke}, using a pessimistic LISA sensitivity curve at high frequencies, rather than  the official sensitivity curve of the LISA proposal used in~\cite{Caputo:2020irr}. 
This highlights the importance of the sensitivity curve at high frequencies, as well as of the other assumptions entering the estimates. Note that the analysis of
\cite{Moore:2019pke} is restricted to binary BHs with merger times up to 100~yr. As discussed in
\cite{Moore:2019pke}, a  numerous population with larger merger times could also be present. For these wide and slowly inspiraling binaries, the signal is closer to monochromatic and data analysis simplifies considerably, allowing for a lower SNR threshold.
Furthermore, once detected the coalescence in ET, one could go back to archival LISA data, using a lower signal-to-noise threshold. Except for the possibility of giving a forewarning of the coalescences at ET and electromagnetic telescopes, such a search into archival LISA data would still preserve all the other advantages of multi-band detection mentioned below.}

Multi-band observations would have many benefits: a joint LISA-ET detection would provide
sky localization of the source with an error of only a few square degrees, and 
would make it possible to alert telescopes and look for an electromagnetic counterpart (which in principle is not expected for BH-BH coalescences, but could be present in BH-NS binaries) both  in the pre-merger and post-merger phases; it
would improve parameter estimation, reducing the error on the luminosity distance to the source and on the initial spins  and  allowing to measure with extreme precision the sky position, mass and spin of the final BH. LISA and ET observations of such events would be highly complementary; for instance LISA, by observing the long inspiral phase, will measure very accurately the masses of the initial BHs, while ET  would detect  the last few cycles and the merger, and would therefore measure the final masses and spin from the ringdown of the final BH. Consistency tests  between the
inspiral part of the waveform and the merger-ringdown part, of the type performed in \cite{TheLIGOScientific:2016src} for the first detection GW150914, would then provide very stringent tests of General Relativity~\cite{Vitale:2016rfr}.
Furthermore, the early warning provided by LISA on particularly interesting events might allow real time optimization of ET  to improve sensitivity to the ringdown signal~\cite{Tso:2018pdv}.

\section{Fundamental physics and cosmology}

\begin{figure}[t]
\centering
\includegraphics[width=0.65\textwidth]{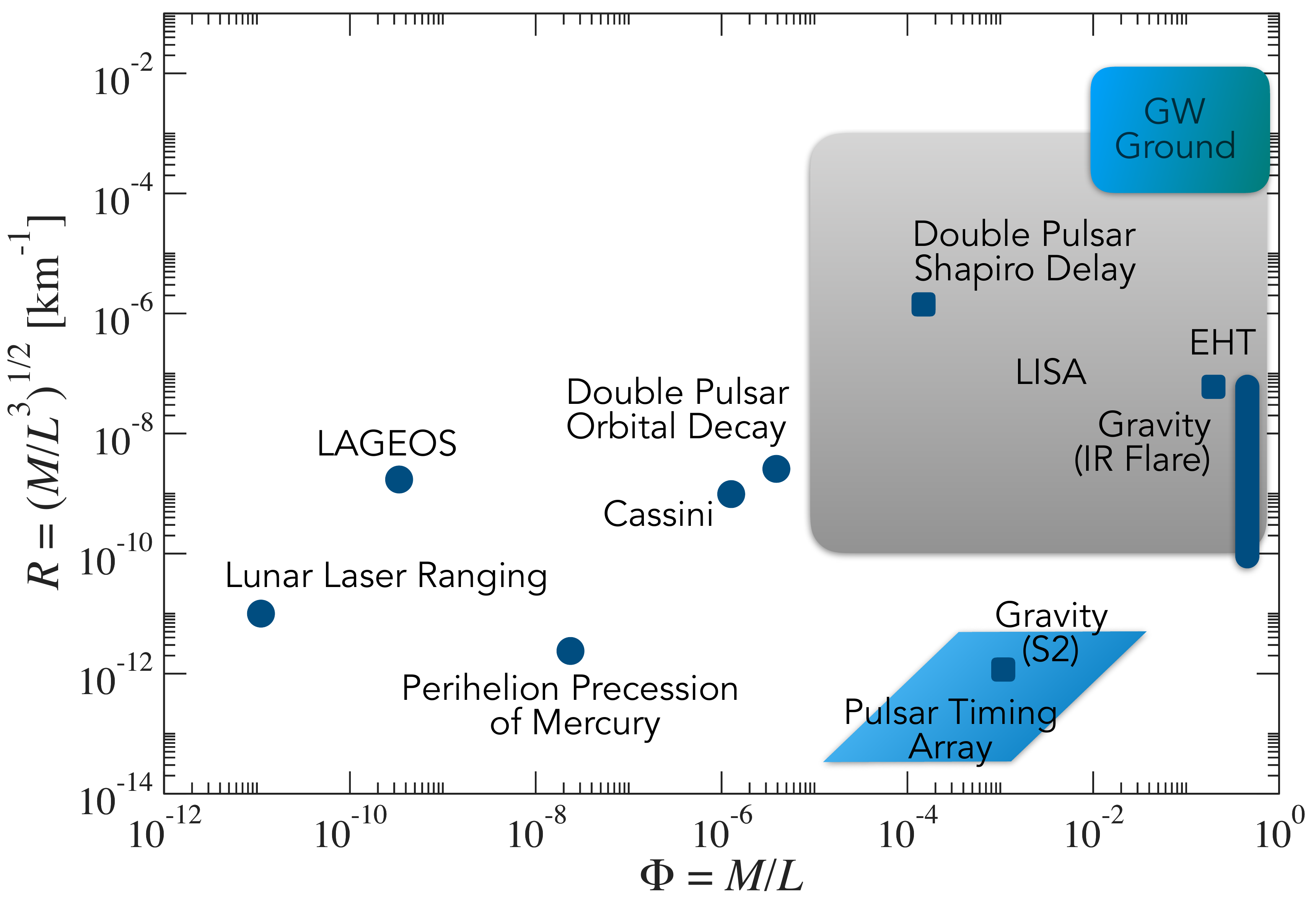}
\caption{Probing gravity at all scales: illustration of the reach in spacetime curvature versus potential
energy targeted by different  kinds of observations. 
$M$ and $L$ are the characteristic mass and length involved in the system or process being observed. 
The genuinely strong-field dynamics of spacetime 
manifests itself in the top right of the diagram. The label EHT refers to the Event Horizon Telescope.
From ref.~\cite{3GScienceBook,Sathyaprakash:2019yqt}.
}
\label{fig:phasediagram}
\end{figure}

The direct detection of gravitational waves has started to give us access to the genuinely
strong-field dynamics of spacetime. This is illustrated in Fig.~\ref{fig:phasediagram}, which shows how 
different kinds of observations (past, current, and future) will give us access to different
regimes, in terms of spacetime curvature $R$ and gravitational potential $\Phi$.\footnote{In a binary system, the gravitational potential felt by a star because of the presence of the companion is of order $v^2/c^2$,
where $v$ is the characteristic orbital velocity of the binary and $c$ the speed of light; therefore, it becomes strong when the system is close to relativistic. However, for compact objects such as neutron stars,  the gravitational potential generated by the star itself in its interior and close to its surface is already of order one. This can induce  strong-field effects on the binary evolution even when the orbital velocity $v\ll c$. This  can happen for instance in some scalar-tensor theories, through the phenomenon of spontaneous and induced scalarization~\cite{Damour:1992we,Damour:1993hw}, as well as 
dynamical scalarization~\cite{Barausse:2012da}. These effects can indeed be constrained by GW observations, see \cite{Sampson:2014qqa} for a study with 2G detectors.}

Observations of GWs from binary BH and binary  NS 
coalescences with Advanced LIGO and Advanced Virgo have enabled us to probe for the first time
the regime where both $R$ and $\Phi$ are no longer small. By observing the inspiral phase  we could test the predictions of GR (as encoded in the post-Newtonian coefficients) to a precision of about $10\%$.
By observing the full  inspiral-merger-ringdown process
of binary black holes, we could perform a first study of the dynamics of vacuum spacetime. 
The observation of the binary neutron star inspiral GW170817 also gave us our empirical 
access to the interaction of spacetime with high-density matter. Because of the large 
distances that GWs have to travel from source to observer, we were able to
strongly constrain possible dispersion that might occur; the latter led to a bound on 
the mass of the graviton of $m_g \leq 5 \times 10^{-23}\,\mbox{eV}/c^2$. This example
notwithstanding, on the whole the existing detectors lack the sensitivity to put very strong
constraints on possible deviations from Einstein's theory, particularly regarding 
the strong-field dynamics at the source, corresponding to the top right edge of 
Fig.~\ref{fig:phasediagram}. 
 
The situation will be quite different with Einstein Telescope. One reason is the much 
larger detection rate; especially for the purposes of fundamental physics, information 
from multiple sources can often be combined, and the measurement accuracy on 
common observables  
tends to improve with the square root of the number of detections. 
For example, the post-Newtonian coefficients that govern binary inspiral will be determined with sub-percent to sub-permille accuracy.
However, the fact that
the same GW source will be much louder in ET will also give us access to 
qualitatively new effects. Below we discuss in turn capabilities of ET in 
probing the properties of gravity, as well as unraveling the nature of ultra-compact
objects, with potentially game-changing implications for our understanding of black holes, 
the make-up of dark matter, dark energy, and maybe even quantum gravity itself.

\subsection{Physics near the black hole horizon: from tests of GR to quantum gravity}

\vspace{2mm}

\subsubsection{Testing the GR predictions for space-time dynamics near the horizon}

Black holes are one of the most extraordinary predictions of General Relativity. They are identified through  their most striking property: in the case of stellar mass black holes, a mass $O(10-100)\Msun$ is concentrated in an extremely small volume; for instance, the Schwarzschild radius of a non-rotating BH with mass $10\msun$ is about 30~km. However, how certain can we be that the massive compact objects that we saw merge with 2G detectors are really the standard black holes of classical General Relativity?

General Relativity gives detailed and specific predictions on the nature of BHs that a 3G detector such as ET will be able to test. The celebrated no-hair theorem  of GR states that, in  a stationary situation, a BH is determined
by just two numbers: its mass and its spin (plus the electric charge, which however is not relevant in an astrophysical context, where it is quickly neutralized). However, when a BH is perturbed, it reacts in a very specific manner, relaxing to its stationary configuration by oscillating in a superpositions of 
quasi-normal modes, which are damped by the emission of GWs.\footnote{The expression `quasi-normal modes', in contrast to `normal modes', emphasizes that the normal modes are unstable to GW emission.} The fact that an elastic body has normal modes is a familiar notion from elementary mechanics. It is however quite fascinating to realize  that a BH, which is a pure space-time configuration, also has its quasi-normal modes. These represents pure space-time oscillations, in a regime of strong gravity, and, in a sense, describe the elasticity of space-time in a most extreme situation, in the region close to the BH horizon. The theory of BH quasi-normal modes is a classic chapter of GR (see \cite{Berti:2009kk,Maggiore:2018zz} for reviews), and in particular predicts the spectrum of frequencies and damping times of the quasi-normal modes as a function of the mass and spin of the BH. Highly perturbed black holes arise as the 
remnants of binary BH or NS mergers, and relax to the final stationary BH configuration through GW emission in the quasi-normal modes, in the so-called `ringdown' phase of the coalescence, where the waveform is given by a superposition of damped sinusoids. Indeed, for the first observed BH-BH coalescence, GW150914, the final ringdown phase was visible, and was shown to be broadly consistent with the prediction of GR for the value of the parameters inferred from  the inspiral part of the waveform~\cite{TheLIGOScientific:2016src}. 

\begin{figure}[t]
\centering
\includegraphics[width=0.65\textwidth]{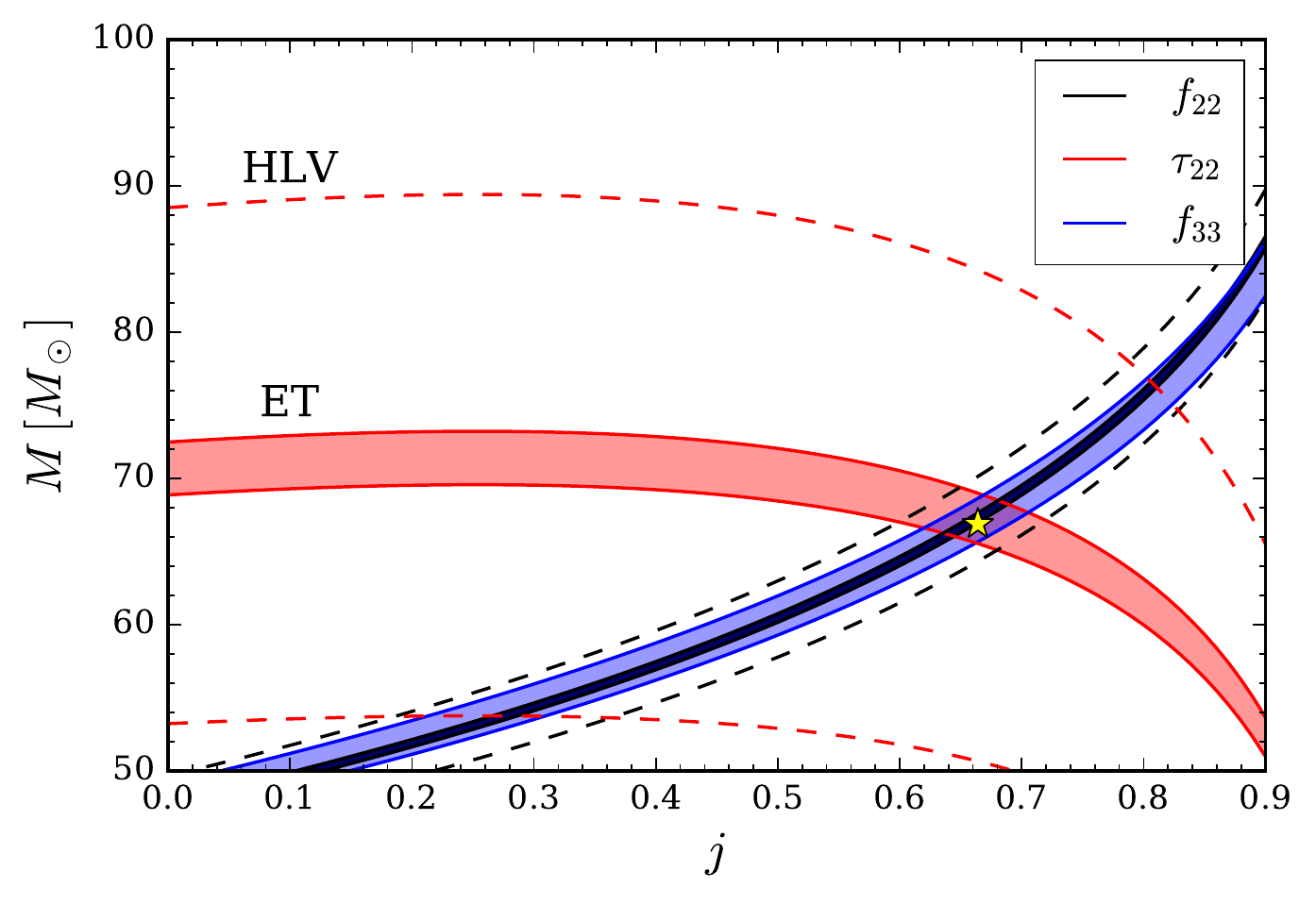}
\caption{Testing the nature of black holes by using two quasi-normal modes and checking
that the characteristic frequencies $f_{22}$ and $f_{33}$ and the damping time $\tau_{22}$ are consistent
with each other, given that for ordinary black holes these can only depend on two numbers, namely
the final mass $M$ and final spin $j$. The estimates are for the ``ringdown" of the remnant
black hole arising from a binary similar to the source of GW150914. The dashed curves marked HLV 
are the 95\% confidence regions one would obtain from Advanced LIGO-Virgo, while the colored bands
are for ET. The star indicates the true values of $M$ and $j$. Figure  from ref.~\cite{3GScienceBook}.
}
\label{fig:ringdown}
\end{figure} 

Since the whole spectrum of frequencies and damping times of the quasi-normal modes depends only on two parameters (the mass $M$ and the spin $j$ of the final BH)  a stringent test of GR can be performed if we can  
measure (at least) three independent quantities. Fig.~\ref{fig:ringdown} illustrates the difference in the accuracy of such a test between 2G and 3G detectors, for a single source such as GW150914 (see \cite{Brito:2018rfr} for the 2G result). Furthermore, 
the accuracy of the measurement scales as $1/\sqrt{N}$, where $N$ is the number of detections;
as we saw, ET will detect $N\sim O(10^5-10^6)$ BH binaries per year,   compared to several hundreds expected for 2G detectors at target sensitivity. Out of these, a large fraction will  have a detectable ringdown signal. In \cite{Berti:2016lat}, 
using three different population synthesis models and the ET-D sensitivity curve, it was found  that ET,   as a single instrument, will detect between $10^4$ and a few times $10^5$ binary BH mergers per year in which the final ringdown phase has a signal-to-noise ratio  larger than~8. If one further requires a ringdown signal sufficiently strong to be able to disentangle the fundamental $l=m=2$ QNM from a QNM with different $l$ (in general, the $l=m=3$ mode, except in the comparable mass limit, where odd-$m$ modes are suppressed, and it becomes easier to resolve the $l=m=4$ mode), one finds that ET will still detect  $20-50$ events per year, with redshifts up to $z\simeq 2$ (see Fig.~2 of \cite{Berti:2016lat}). For comparison, in 2G detectors the  rate  for resolving two different QNMs would be of order $10^{-2}$ events per year~\cite{Berti:2016lat}. ``Black hole spectroscopy'' is therefore out of reach for 2G detector, while it will be routinely performed with 3G detectors.
In the comparable mass limit, where the $l=m=3$ mode is suppressed, another option is to resolve some overtones of the $l=m=2$ mode. As shown in~\cite{Bhagwat:2019dtm}, this requires a ringdown signal-to-noise ratio larger than $\sim 30$, which will be routinely achieved by ET. 

One can go beyond consistency tests of GR either by working out explicit predictions for the quasi-normal mode frequencies and damping time as a function of $(M,j)$ in specific extensions of GR, or by introducing parametrizations of deviations from the Kerr predictions (in the same spirit as the parametrization of deviations from the post-Newtonian predictions expressed by the parametrized Post-Newtonian expansion).  Using the parametrization  developed in \cite{Maselli:2019mjd} it is found that $O(10)$ ringdown detections at a signal-to-noise ratio $\sim 100$, as can be obtained at ET, can provide significant constraints on the first  `beyond-Kerr' parameters.

\subsubsection{Exotic compact objects and signals from quantum gravity}

The observation of quasi-normal modes, beside providing a spectacular test of GR in the strong-field, near-horizon regime, could also potentially lead to the discovery of different types of compact objects.
Indeed, various exotic compact objects have been proposed that may act as ``black hole mimickers",
such as boson stars, gravastars, stars composed of dark matter particles, etc.
(see \cite{Cardoso:2019rvt} for review). When such  objects are part of a binary system that undergoes
coalescence, they can make their presence known through various possible imprints on 
the GW signal emitted. Already during the inspiral phase, these objects may get tidally deformed in a way that would be impossible 
for a standard, classical black hole. Unlike second generation detectors, ET will for instance be able to distinguish neutron stars from boson stars even for the most compact models of the latter~\cite{Cardoso:2017cfl}.  
Another possibility is that an exotic object could be identified through 
an anomalous spin-induced quadrupole moment, which would again not be accessible with 
current detectors, but measurable with ET to the percent level~\cite{Krishnendu:2017shb}.

If the outcome of a coalescence is different from a BH, this might leave an imprint on  the  ringdown phase, and could be tested by measuring quasi-normal mode frequencies and life-times, as in Fig.~\ref{fig:ringdown}. For exotic compact objects where the modifications take place only at  scales much shorter than the so-called light-ring (as in the case of quantum gravity effects discussed below), the ringdown signal will be very similar to a BH, but after the ringdown has died down, exotic compact objects may continue to 
emit bursts of gravitational waves at regular time intervals, called \emph{echoes}~\cite{Cardoso:2016rao,Cardoso:2017cqb}. The detectability of echoes  has been investigated in~\cite{Maggio:2019zyv,Testa:2018bzd}, that show that excluding or detecting echoes   requires signal-to-noise ratios in the post-merger phase of $O(100)$, that  will be achievable only with 3G detectors such as ET and CE, or with LISA.


Prompted by Hawking's 
information paradox, modifications of the structure of space-time at the horizon scale have been proposed, such as firewalls~\cite{Almheiri:2012rt} and fuzzballs~\cite{Mathur:2005zp}, for which the classical horizon is removed through macroscopic quantum effects. The absence of the horizon results in an instability, that leads to a detectable stochastic background of gravitational waves~\cite{Barausse:2018vdb}.
From a particle physics perspective, one is used to the fact that, at energies $E$ much below the Planck energy scale $M_{\rm Pl}$, quantum gravity effects are suppressed by powers of 
$E/M_{\rm Pl}$, and therefore, given that  the Planck scale $M_{\rm Pl}$ is of order $10^{19}$~GeV, they are totally unaccessible at accelerators, even in any foreseeable future. Equivalently, at a macroscopic length-scale $L$, quantum gravity effects are suppressed by powers of $l_{\rm Pl}/L$, where $l_{\rm Pl}\sim 10^{-33}$~cm is the Planck length. In contrast, near the BH horizon, where the characteristic length-scale $L$ is given by the Schwarzschild radius $R_S$, effects due to quantum gravity  are governed  by a factor  $\log (l_{\rm Pl}/R_S)$, and can manifest themselves  through a series of echos after the initial ringdown signal~\cite{Cardoso:2016rao,Cardoso:2016oxy,Cardoso:2019apo}, emitted with a time delay  $\tau_{\rm echo}\sim (R_S/c) \log (R_S/l_{\rm Pl})$. For instance, for a final object with mass $M=60\Msun$, one has 
$\tau_{\rm echo}\simeq 16\tau_{\rm BH}$, where $\tau_{\rm BH}\simeq 3$~ms is the fundamental damping time of a Schwarzschild BH with this mass. Such signals 
 are potentially within the reach of ET, which raises the tantalizing possibility of accessing 
quantum gravity effects at ET.

\vspace{1mm}
To summarize, {\em the transition from second generation observatories to Einstein Telescope
will lead to a qualitative leap in our ability to probe both the nature of gravity in the strong field regime 
and the structure of compact objects, and could even lead to exploring the quantum gravity regime.}

\subsection{The nature of dark matter}

From cosmological observations we know that the elementary constituents that we observe everyday in the lab (electrons, protons, atoms, etc.) are responsible for only about $5\%$ of the energy density of the Universe; about $25\%$ is made of ``dark matter", i.e. matter that does not have electromagnetic interactions and whose existence is only inferred through its gravitational attraction, while the rest, about $70\%$, is currently attributed to ``dark energy", a form of energy that formally produces a negative pressure and therefore cannot be identified with any known or unknown type of matter. Understanding the nature of dark matter and of dark energy is one of the crucial problems in astrophysics, cosmology and fundamental physics.  ET will be able to  shed light on both questions. In this section we discuss the potential of ET for dark matter studies, while 
its potential for dark energy will be discussed in Sect.~\ref{sec:cosmos}.

Observations at ET will allow us to attack the problem of the origin of dark matter from several different angles. Dark matter could be composed, at least in part, of \emph{primordial} black holes in the mass range $\sim 0.1 - 100\,M_\odot$. As we already mentioned in Section~\ref{sect:Blackholebinaries},
primordial BHs  could be seeded by fluctuations generated during the 
last stages of inflation, which then collapsed in later epochs as a consequence of drops in the pressure of the cosmic fluid, e.g. during the  QCD quark-hadron transition. Their mass distribution depends on the precise model 
of inflation and on the epoch when they  collapsed. The large number of mergers that Einstein Telescope
will see, together with its  ability to access a broad range of masses, would allow us to map the black hole mass distribution
and identify an excess of black holes in certain mass intervals. For black holes with masses well below a 
solar mass, no plausible astrophysical formation mechanism is available, so that their detection would
point to the existence of primordial black holes. A unique advantage of Einstein Telescope is the possibility of observing stellar-mass black hole mergers at redshifts of $\sim 10-20$, before
any stars had formed that could create black holes in the usual way; 
should such an event be observed then (irrespective of masses) 
the objects involved are bound to be of primordial origin.

If most of the dark matter occurs in the form of particles beyond the Standard Model, 
then also in that case gravitational wave observations can be used to find them. Black holes
could not only accrete dark matter particles, but also be subject to gravitational drag, which 
in a binary system would accumulate over the course of many orbits. If Einstein 
Telescope will be operational during the same period as LISA, joint LISA-ET observations
of the same source will then be of great value~\cite{Barausse:2016eii}.

There is also the possibility that dark matter particles are captured in astrophysical
objects and thermalize with the star~\cite{Gould:1989gw}. The presence of a dark matter core in a neutron star
might again have an imprint upon the GW signal during binary inspiral and merger.
Dark matter accumulating in neutron stars and interacting through Yukawa-like interactions in the dark sector could affect the orbital dynamics of a neutron star binaries, and therefore the corresponding waveform, in a way detectable by ET~\cite{Alexander:2018qzg}, whose low-frequency sensitivity makes it an especially sensitive probe to dark matter mediated forces between neutron stars.
In some models~\cite{Bramante:2017ulk,Kouvaris:2018wnh}, the accumulation of dark matter may lead to the formation of a black hole 
inside a neutron star, which then accretes the remaining neutron star matter, leading to 
black holes of $(1-2) \,M_\odot$ that could be observed by ET. 

Finally, ultralight bosons have been proposed in various extensions of the Standard Model, and also as  dark matter candidates~\cite{Essig:2013lka,Hui:2016ltb}. If their
Compton wavelength is comparable to the horizon size of a stellar or supermassive 
rotating black hole (\emph{i.e.}~for particle masses of $10^{-21} - 10^{-11}$ eV),
they can extract rotational kinetic energy from the black hole through ``superradiance" 
to feed the formation
of a bosonic ``cloud" with mass up to $\sim 10$\% of the black hole~\cite{Arvanitaki:2010sy,Brito:2014wla, East:2017ovw}. These clouds
annihilate over a much longer timescale than their formation, through the emission of 
nearly monochromatic gravitational waves which could be detected either directly 
or as a stochastic background from a large number of such objects throughout the Universe~\cite{Brito:2017zvb, Brito:2017wnc}.
Additionally, measuring the distribution of black hole masses and spins can yield
an indication of the prevalence of superradiance through light scalars. Moreover, 
the presence of such clouds will again have an effect on binary orbital motion~\cite{Baumann:2018vus}. 
This way gravitational waves have the potential to provide a unique probe into 
an ultralight, weakly coupled regime of particle physics that can not easily be
accessed in accelerator experiments.

{\em To summarize, ET has the potential of discovering several dark-matter candidates that will be unaccessible by any other means.}

\subsection{The nature of dark energy}\label{sec:cosmos}

ET will be an outstanding discovery machine for studying the nature of dark energy, using binary NSs and binary BHs as cosmological probes.
Indeed, a remarkable feature of the GWs emitted in the coalescence of compact binaries is that  their signal   provides an absolute measurement of the luminosity distance to the source. The relation between the luminosity distance $d_L$ and redshift $z$ of the source carries crucial cosmological information and is among the main observables of modern cosmology. Explicitly, it is given by
\begin{equation}\label{eq:dL(z)}
d_L(z)=\frac{1+z}{H_0}\int_0^z \frac{dz'}{\sqrt{\Omega_M(1+z')^3+\frac{\rho_{\rm DE}(z')}{\rho_0}}},
\end{equation}
where $H_0$ is the Hubble parameter, $\rho_0$ is the closure energy density, $\rho_{\rm DE}$ is the dark energy density and $\Omega_M=\rho_M(t_0)/\rho_0$ is  the  density  of matter at the present time $t_0$, normalized to $\rho_0$ (and we neglected for simplicity the contribution of radiation, which is irrelevant at the redshifts of interest for GW detectors, and a possible non-vanishing spatial curvature). In particular, in  $\Lambda$CDM, which is the model that constitutes the  current cosmological paradigm, $\rho_{\rm DE}(z)/\rho_0=\Omega_{\Lambda}$ is a constant, related to the cosmological constant.

Observations performed with electromagnetic waves can  infer the redshift of a source,  through spectroscopic or photometric observations; however, obtaining the absolute distance to a source at cosmological distances is much more difficult. Ideally, this requires the existence of a ``standard candle'', a class of sources whose intrinsic luminosity ${\cal L}$ is known, so that, from a measurement of the  energy flux ${\cal F}$  received by the observer, we can reconstruct the luminosity distance $d_L$ from ${\cal F}={\cal L}/(4\pi d_L^2)$.
A classic example of standard candle in cosmology is provided by type~Ia supernovae: these are bright enough to be visible at cosmological distances, and, after some empirical corrections, their intrinsic luminosity can be considered as fixed; its value is then calibrated through the
 construction of a  ``cosmic distance ladder'', in which classes of sources at shorter distances are used to calibrate different sources at higher and higher distances. Indeed, type Ia Supernovae provided the first conclusive evidence for the  existence of dark energy~\cite{Riess:1998cb,Perlmutter:1998np}, a discovery that was awarded with the 2011 Nobel Prize in Physics.
 
GW observations of compact binary coalescences completely bypass the need for empirical corrections and the uncertainties in the  calibration of the cosmic distance ladder, since the observed waveform of the inspiral phase directly carries  the information on the luminosity distance $d_L$~\cite{Schutz:1986gp}. In this context, coalescing binaries are called ``standard sirens'', the GW analogue of standard candles. By contrast, the GW signal does not carry  direct information on the redshift, so the situation is reversed compared to electromagnetic observations. 
The ideal situation then takes place when one has a joint GW-electromagnetic detection, as was the case for the NS-NS binary GW170817. In this case the GW signal gives $d_L$ and the electromagnetic observation  the redshift $z$. Even in the absence of a redshift determination from an electromagnetic counterpart, several statistical methods have been discussed in the literature, to extract cosmological information from purely GW observations.\footnote{For instance, in a modern version of a statistical method proposed already in \cite{Schutz:1986gp}, one  constructs a Bayesian likelihood by assigning probabilistically a host galaxy to the GW event, within the localization volume found by the GW detectors~ (see \cite{DelPozzo:2011yh} for a recent discussion).  After collecting a sufficient number of events, the results will converge to the actual $d_L-z$ relation. Other techniques, applicable to NSs, exploit the narrowness of their intrinsic mass distribution~\cite{Taylor:2011fs,Taylor:2012db} or
the effect of the equation of state in the inspiral phase~\cite{Messenger:2011gi}.}

In the low redshift  limit $z\ll  1$ accessible to 2G detectors, eq.~(\ref{eq:dL(z)}) reduces to the Hubble law $d_{L}(z)\simeq H^{-1}_0z$. Hence the observation of standard sirens  at low redshifts can provide  a measurement of  $H_0$, but is insensitive to the dark energy density $\rho_{\rm DE}$, or equivalently to the dark energy equation of state $w_{\rm DE}$. The possibility of measuring $H_0$  has already been demonstrated with  GW170817, from which a value  $H_0=70.0^{+12.0}_{-8.0}\,\, {\rm km}\, {\rm s}^{-1}\, {\rm Mpc}^{-1}$ was obtained~\cite{Abbott:2017xzu}.  With $O(100)$ standard sirens with counterpart, or with statistical methods, a measurement of $H_0$ at the $1\%$ level could already be possible with 2G detectors. This would  allow us to  arbitrate the current discrepancy between the  value of the Hubble parameter $H_0$ obtained from  late-Universe probes~\cite{Riess:2019cxk,Wong:2019kwg}, and the  value  inferred from early-Universe probes~\cite{Aghanim:2018eyx,Abbott:2018xao}, which has currently reached the  $5.3\sigma$ level.

\begin{figure}[t]
\includegraphics[width=0.48\textwidth]{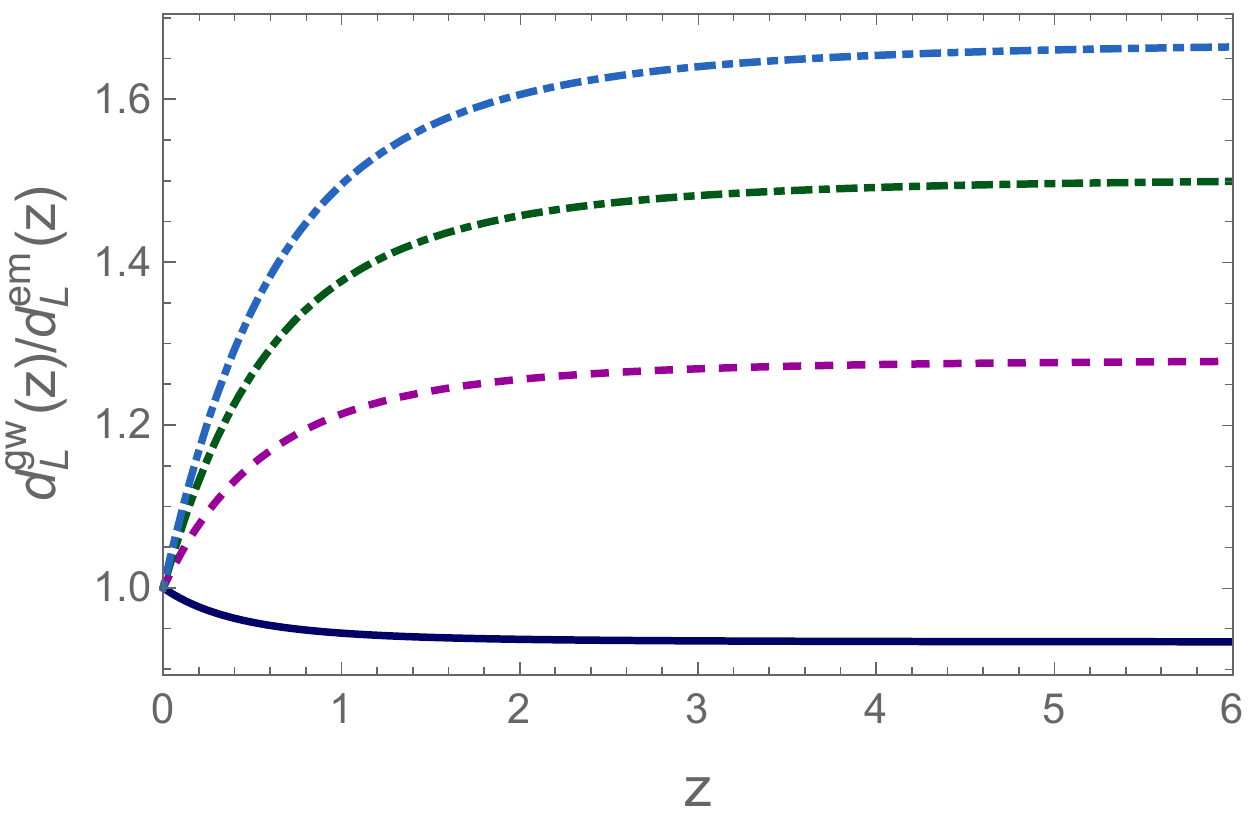}\quad\quad
\includegraphics[width=0.34\textwidth]{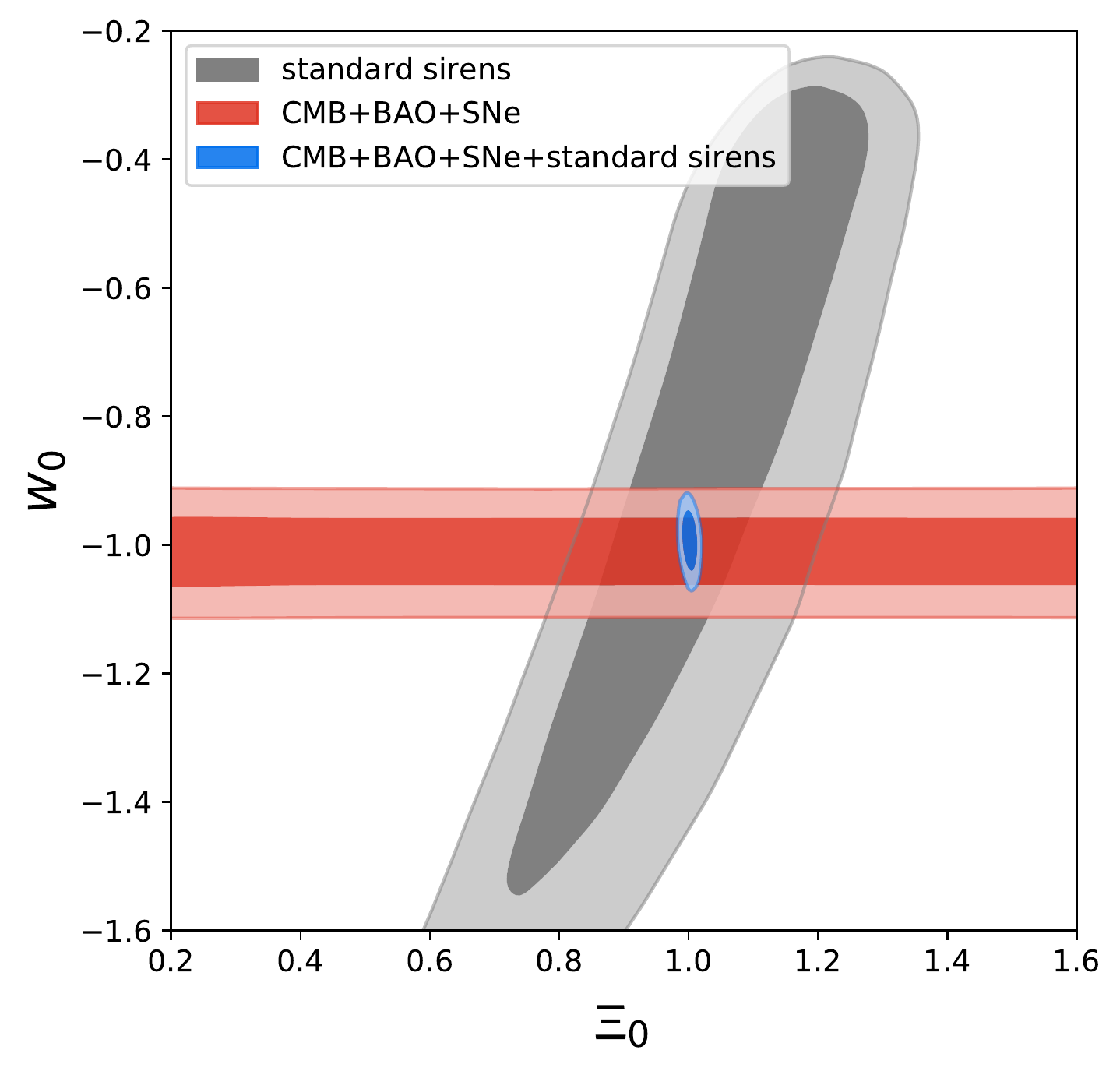}
\caption{Left panel: an example of the ratio of the gravitational to electromagnetic luminosity distance in a modified gravity model, for different values of a parameter of the model. In the upper curve, at large redshift  the deviations of $d_L^{\rm gw}(z)$ from $d_L^{\rm em}(z)$ reaches $60\%$ (from 
ref.~\cite{Belgacem:2019lwx}). Right panel: constraints on the parameters $(\Xi_0,w_0)$  that describe a non-trivial dark energy sector. Modified GW propagation is not accessible by electromagnetic observations from CMB, Baryon Acoustic Oscillations,  and Supernovae, whose contours (red) are flat along the $\Xi_0$ direction. Standard sirens at ET (gray), combined with these electromagnetic probes  allow a determination of $\Xi_0$ 
 to better than $1\%$ (blue). (From ref.~\cite{Belgacem:2018lbp}).
\label{fig:xi0w0}}
\end{figure}

At ET, given  the expected huge number of detections and the very high signal-to-noise ratios  of nearby  events, a sub-percent level accuracy on $H_0$  could be reached. However,
a much higher potential for discovery is provided by the fact
that ET will have access to standard sirens at much larger redshifts, where effects genuinely induced by a non-trivial  dark energy  sector and by modifications of General Relativity on cosmological scales  become accessible. First of all, away from the limit $z\ll 1$, the effect of  the dark energy 
density $\rho_{\rm DE}(z)$  becomes visible in eq.~(\ref{eq:dL(z)}), which would allow us to obtain a   measurement of the dark energy equation of state from GW observations~\cite{Sathyaprakash:2009xt,Zhao:2010sz,Belgacem:2018lbp}. In fact, the situation for 3G detectors is even more interesting due to a phenomenon of modified GW propagation~\cite{Saltas:2014dha,Lombriser:2015sxa,Nishizawa:2017nef,Belgacem:2017ihm,Amendola:2017ovw,Belgacem:2018lbp,Belgacem:2019pkk}. Indeed, a natural theoretical framework for having a  dark energy sector different from a simple cosmological constant is provided by modifications of GR at the cosmological scale. In a generic modified gravity theory the cosmological evolution of the background is different from that of $\Lambda$CDM, and this is encoded in a non-trivial dark-energy density $\rho_{\rm DE}(z)$
[or, equivalently, in the dark energy equation of state $w_{\rm DE}(z)$)]. On top of this, cosmological perturbations will also be different. The modification in the scalar perturbation sector  will affect the predictions for the growth of structures or lensing, and
are among the  targets of future experiments such as Euclid, DESI or SKA. The modification in the tensor perturbation sector will instead affect the propagation of GWs over cosmological distances. In GR the GW amplitude scales as the inverse of the scale factor, $h\propto 1/a$, which eventually results in the fact that the signal from coalescing binaries at cosmological distances is proportional to $1/d_L(z)$. In  modified gravity theories (including the models where GWs propagate at the speed of light, so to comply with the limit imposed by GW170817), this behavior is changed.\footnote{In GR the equation of propagation of GWs over a FRW background is given by $\tilde{h}''_A  +2 {\cal H}\tilde{h}'_A+k^2\tilde{h}_A=0$, where $h_A$ is the GW amplitude, $A=+,\times$ labels the polarization, and ${\cal H}=a'/a$ is the Hubble parameter. In alternative theories this is modified to
$\tilde{h}''_A  +2 {\cal H}[1-\delta(\eta)] \tilde{h}'_A+k^2\tilde{h}_A=0$, where the function $\delta$ encodes the details of the modification.}
As a result, the GW amplitude becomes inversely proportional to a ``GW luminosity distance'', different from the standard electromagnetic one. An example of the  resulting ratio of GW luminosity distance to the standard `electromagnetic' luminosity distance, in a specific model of modified gravity, is shown in the left panel of Fig.~\ref{fig:xi0w0}.
This behavior turns out to be  completely generic to  modified gravity models (scalar-tensor theories, nonlocal modifications of gravity, bigravity, etc.)~\cite{Belgacem:2019pkk}. For most models, the deviations from GR can be parametrized in terms of two parameters $(\Xi_0,n)$  as~\cite{Belgacem:2018lbp}
\begin{equation}
\frac{d_L^{\rm gw}(z)}{d_L^{\rm em}(z)}=\Xi_0+\frac{1-\Xi_0}{(1+z)^n},
\end{equation}
where $d_L^{\rm em}$ is the standard electromagnetic luminosity distance. Measuring the modified GW propagation through its effect on the GW luminosity distance is a very powerful probe for the dark energy sector which cannot be accessed at all with  electromagnetic  observations. With a few hundreds standard sirens with counterpart, ET will constrain $\Xi_0$ to below 1\% (see the right panel in Fig.~\ref{fig:xi0w0}), a level significantly  smaller than the deviation from GR foreseen by various alternative gravity theories. Indeed, the sector of tensor perturbations over a cosmological background  can only be explored with  GW detectors, and can lead to significant surprises. For instance, one can have a cosmological model that is  observationally indistinguishable from 
$\Lambda$CDM in terms of current electromagnetic observations,
 but still, as shown in the left panel of  Fig.~\ref{fig:xi0w0}, predicts a value of $\Xi_0$  that can be as large as $\Xi_0\simeq 1.6$, representing a $60\%$ deviation from $\Lambda$CDM~\cite{Belgacem:2019lwx} (and in fact even up to $80\%$ \cite{Belgacem:2020pdz}).
Such a large effect could be detectable even with just a single standard siren at ET.

\begin{figure}[t]
\centering
\includegraphics[width=0.65\columnwidth]{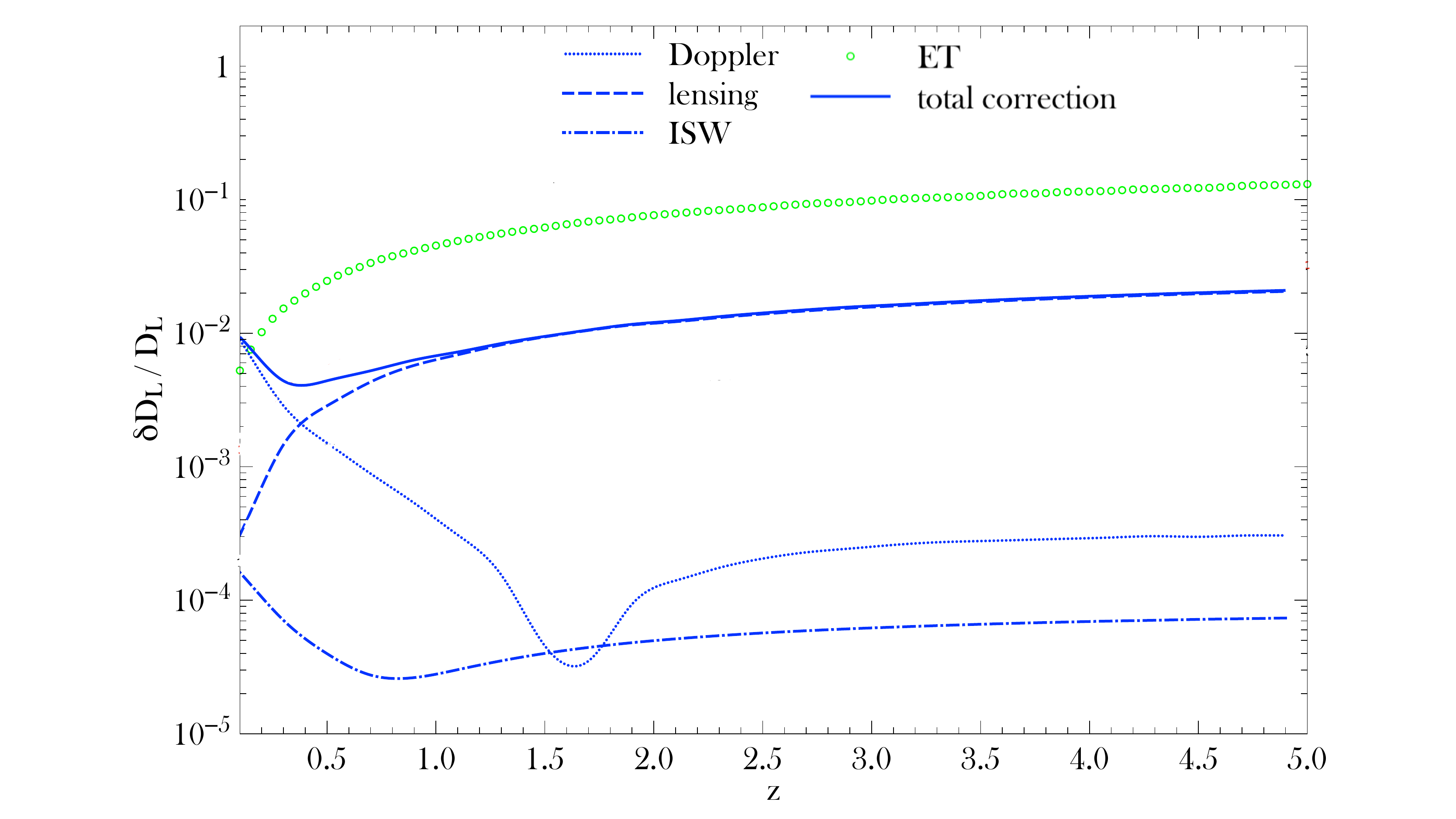}
\caption{Total correction to luminosity distance estimates due to perturbations, as a function of $z$. The dotted, dashed and dot-dashed lines show Doppler, lensing and ISW-like contributions, respectively, while the solid line shows the total effect. The green
points show the predicted precision in measurements of the luminosity distance of ET. Adapted from \cite{Bertacca:2017vod}.}
\label{fig:final}
\end{figure}

{\em In summary, the sector of cosmological tensor perturbations is  virgin territory that can only be explored by third-generation GW detectors such as ET, and which could  offer the most powerful window for understanding the nature of dark energy and modifications of General Relativity at cosmological scales.}

To perform precision cosmography with ET it is also important to investigate 
the effect of cosmological perturbations on the propagation of gravitational waves. 
A detailed analysis  of  the effect of cosmological perturbations and inhomogeneities on estimates of the luminosity distance of BH or NS binary mergers through gravitational waves has been performed
in \cite{Bertacca:2017vod}, where has been  derived an expression 
for the effect of large-scale structures on GW waveforms accounting for lensing, Sachs-Wolfe, integrated Sachs-Wolfe, time delay and volume distortion effects. In Figure~\ref{fig:final} we show the correction  $\delta D_L /D_L$ to luminosity distance estimates due to perturbations, as a function of $z$. The dotted, dashed and dot-dashed lines show velocity, lensing and ISW-like contributions, respectively, while the solid line shows the total effect. The
green points show the predicted precision in measurements of the luminosity distance, at any redshift, for the Einstein Telescope. We see that the additional  uncertainty $\delta D_L/D_L$ due to the inclusion of perturbations is below the error obtained by the sensitivity curve of ET (except at very low $z$ where it becomes comparable) and therefore does not spoil the accuracy that can be obtained by ET.

\subsection{Toward the big bang: stochastic backgrounds of GWs}

The weakness of the gravitational interaction, which is responsible for the fact that  GW detection is such a challenging enterprise, 
also implies that the observed GW signals  carry uncorrupted information about their production mechanism. This is particularly significant for stochastic backgrounds of GWs of cosmological origin. For comparison, in the early Universe photons were kept in equilibrium with the primordial plasma by the electromagnetic interaction, and decoupled from it only at a redshift $z\simeq 1090$, when the Universe already had a rather low temperature $T\simeq 0.26$~eV. The photons that we observe today from the cosmic microwave background  therefore give a snapshot of the Universe at this decoupling epoch, while all information about earlier epochs was obliterated by the photon collisions with the primordial plasma. Neutrinos, which interact through weak interactions, decoupled  when the Universe had a temperature $T\simeq 1$~MeV.  By contrast, GWs were decoupled from the primordial plasma at all temperatures below the Planck scale $\sim 10^{19}$~GeV,
corresponding to a far earlier epoch, and energies far exceeding  those accessible to particle accelerators. {\em The detection of a stochastic background of GWs of cosmological origin would literally provide us with an uncorrupted  snapshot of the earliest moments after the big bang, which would be impossible to obtain by any other probe.}

Stochastic GW backgrounds are characterized by the  energy spectrum $\Omega_{\rm GW}(f)$, which measures the GW energy density per logarithmic interval of frequency, normalized to the critical energy density for closing the Universe; by the angular spectrum, measuring the energy density at different angular scales in the sky; and by their polarization content. In order to detect a stochastic background one has to perform cross-correlation among the outputs of pairs of detectors, as would be possible with a single ET observatory, which is made of three non-parallel detectors.

\subsubsection{Cosmological backgrounds}

One of the targets of ET is the detection and characterization of the stochastic GW background from astrophysical and cosmological sources. On the cosmological side, while the background  generated  by the amplification of quantum vacuum fluctuations due to the inflationary expansion is expected to be too low to be detected by  3G detectors, there are several other inflation-related mechanisms that can produce detectable signals~\cite{Maggiore:1999vm, Caprini:2018mtu, Maggiore:2018zz}. For example, 
large GW amplitudes are naturally produced in inflationary models where there are secondary
fields~\cite{Cook:2011hg} (not responsible for the inflationary period) with arbitrary
spin, and coupled to the inflaton either directly or only through a gravitational coupling, or in models where some symmetries are relaxed during the inflationary period~\cite{Bartolo:2015qvr}. On the other hand, also scenarios alternative to inflation, like e.g. pre-big-bang models inspired by string theory~\cite{Brustein:1995ah,Buonanno:1996xc, Gasperini:2002bn}, predict a spectrum which grows with frequency, resulting in a potentially detectable signal in the ET bandwidth.

\begin{figure}[t!]
\begin{center}
\includegraphics[width=0.6\textwidth]{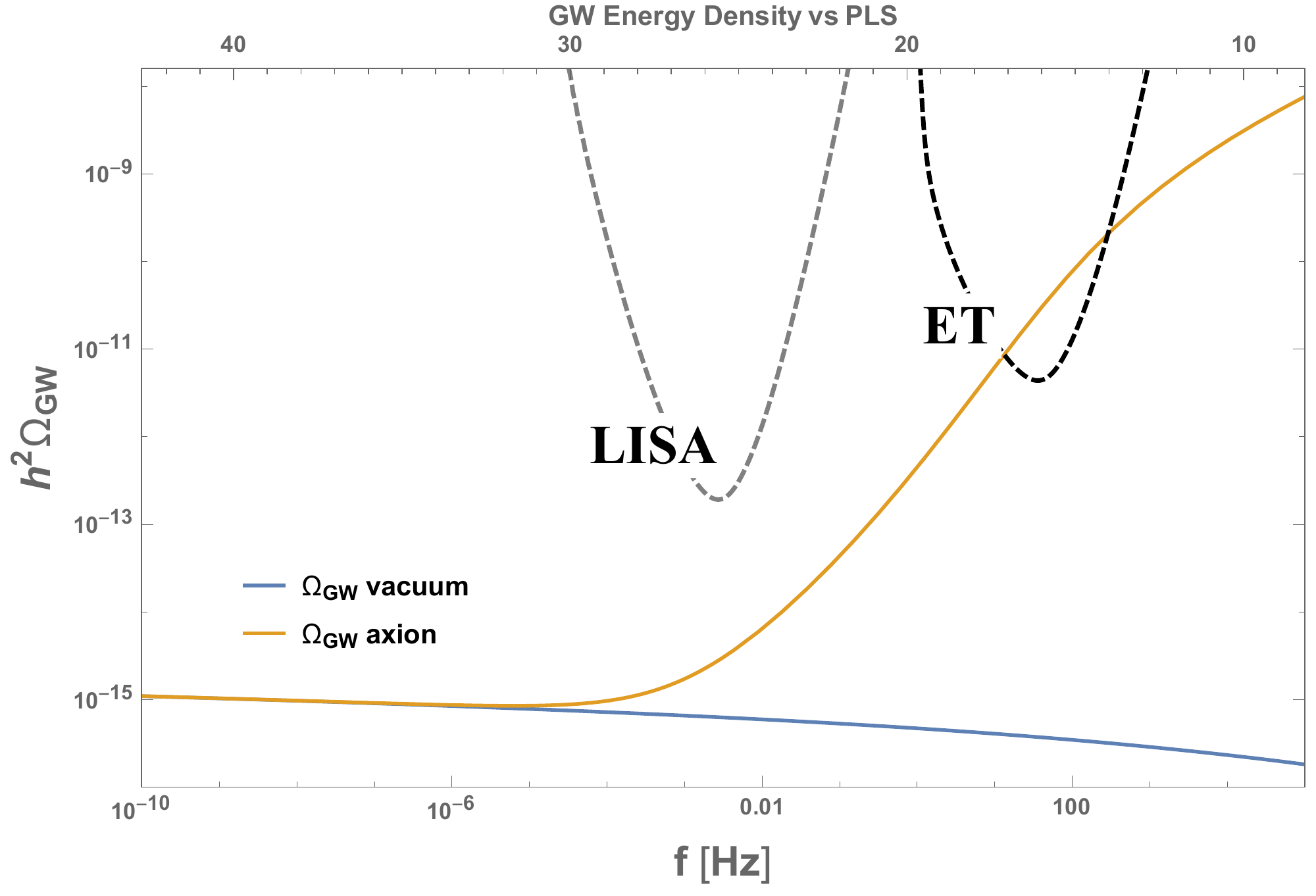}
\end{center}
\caption{Spectrum of GWs today, $h^2 \Omega_{GW}$, for a model of quadratic inflaton potential, with inflaton - gauge field coupling $f = M_{\rm Pl}/34$ vs ET and LISA Power Law Sensitivity (PLS) curves.}
\label{fig:omegaaxion}
\end{figure}

Such cosmological backgrounds, beside having a large spectral energy spectrum $\Omega_{\rm GW}(f)$, can be characterized by peculiar features which ET will have the capability to characterize: for instance, models where the inflaton is coupled to an axion field 
result in a spectrum which increases with frequency (hence, is potentially visible at ET scales while still complying with CMB limits at very low frequencies, see 
Fig.~\ref{fig:omegaaxion}) and is {\it chiral}, with an overproduction of one circular GW polarization with respect to  the other.  ET,  using the dipolar modulation generated by the solar system motion, will be sensitive to such a feature~\cite{Domcke:2019zls}, which is a clear indication of the cosmological origin of the signal.  
Another source of GWs is expected during the (p)reheating period of the Universe, following closely the end of
inflation~\cite{Kofman:1994rk,Kofman:1997yn,Greene:1998nh,Greene:2000ew}. In particular, when ``preheat'' fields are coupled to the inflaton, these may undergo
a non-perturbative excitation after inflation with the consequent generation of GWs. The amplitude of these backgrounds
can be very large, and there are scenarios that can peak at frequencies in the ET range. 3G detectors will also have  the ability to probe post-inflation expansion scenarios where the equation state parameter is {\it stiff},  $1/3 < w\leq 1$~\cite{Figueroa:2019paj}. 
	
Different cosmological sources are expected to have different spectral shapes; however, if one is faced with a superposition of cosmological backgrounds, other observables need to be considered. With a network of 3G detectors the resulting angular resolution, and the possibility of cross-correlating them, will allow us to reach the desired resolution to detect angular anisotropies in the GW energy density. At the same time, the statistical properties (and, particularly, its deviation from gaussianity) of the cosmological stochastic GW background will be another possible target for 3G detectors that will allow to distinguish a cosmological background from other stochastic signals \cite{Bartolo:2019oiq}.

The direct observation of a stochastic GW background with ET will also provide a unique opportunity to test General Relativity during the early universe, through the existence of extra polarization modes of GWs. In GR, a GW has two polarization modes, while in a general metric theory of gravitation,  GWs can have  up to  six independent polarizations. If additional polarizations would be found, it would show that the theory of gravity should be extended beyond GR, and would help to discriminate among theoretical models, depending on which polarization modes are detected.

First-order phase transitions are another potential source of  a stochastic background.
As the Universe expands, its temperature drops and it may undergo a series of phase transitions followed by spontaneous breaking of symmetries. If a phase transition is of first order, 
a  stochastic GW background may be produced as true vacuum bubbles collide and convert the entire Universe to the symmetry-broken phase.
In the Standard Model  of particle physics, the electroweak and the QCD  transitions are just  cross-overs, hence any generated gravitational wave signal is not expected to be detectable. However, there are many extensions of the Standard Model (e.g., with additional scalar singlet or doublet, spontaneously broken conformal symmetry, or phase transitions in a hidden sector) which predict strong first-order phase transitions, not necessarily tied to either the electroweak or the QCD phase transition. In such models, the power of the generated gravitational wave signal depends on the energy available for conversion to shear stress, which is determined by the underlying particle physics model. 
Hence, a stochastic background of GWs will allow us to test particle physics models of the very early Universe, at energy scales far above those  that can be reached at the Large Hadron Collider. 
First-order phase transitions can also lead to turbulence that may generate a stochastic background of gravitational waves.

\begin{figure}[t]
\includegraphics[width=0.45\textwidth]{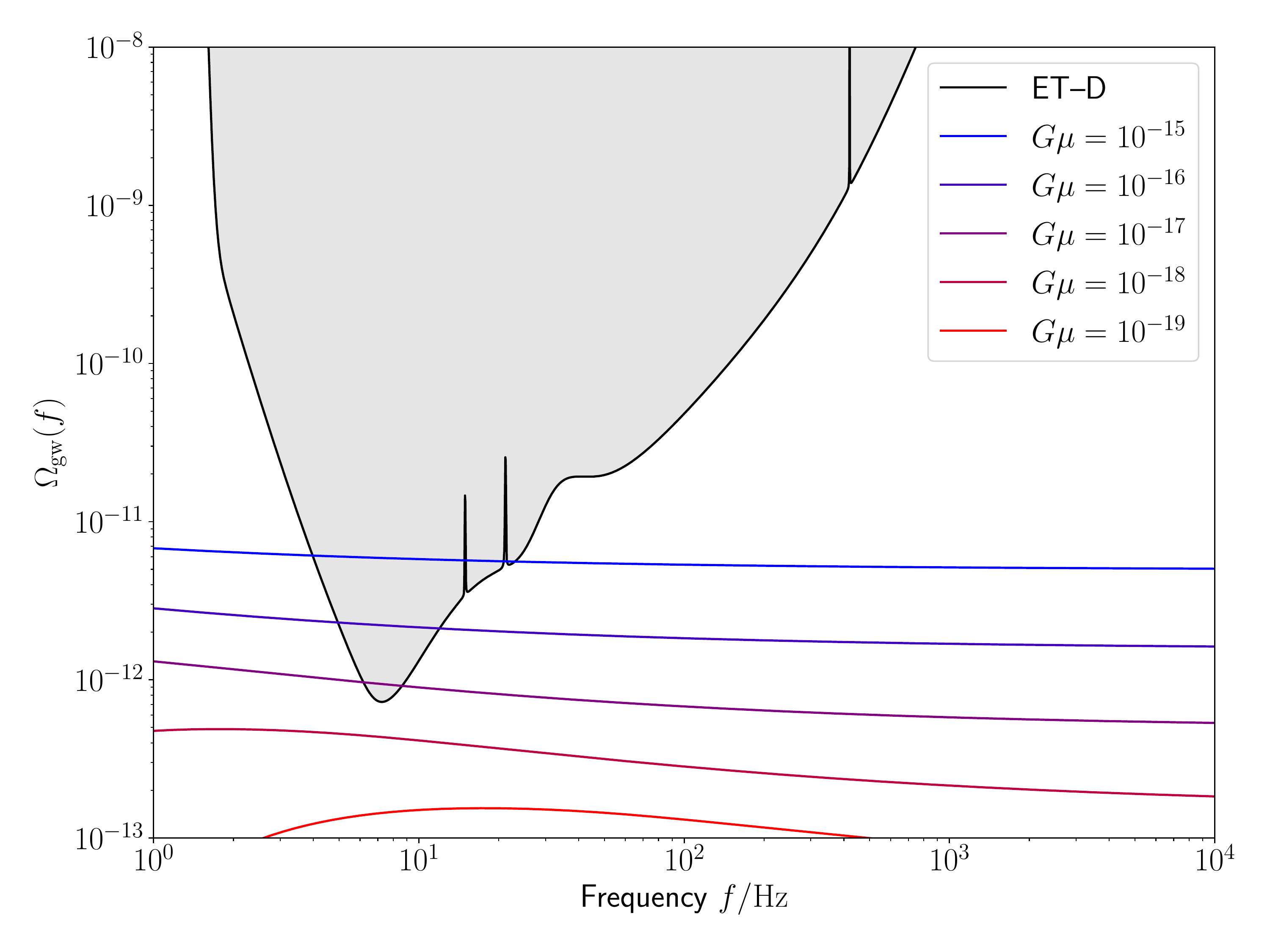}\quad
\includegraphics[width=0.45\textwidth]{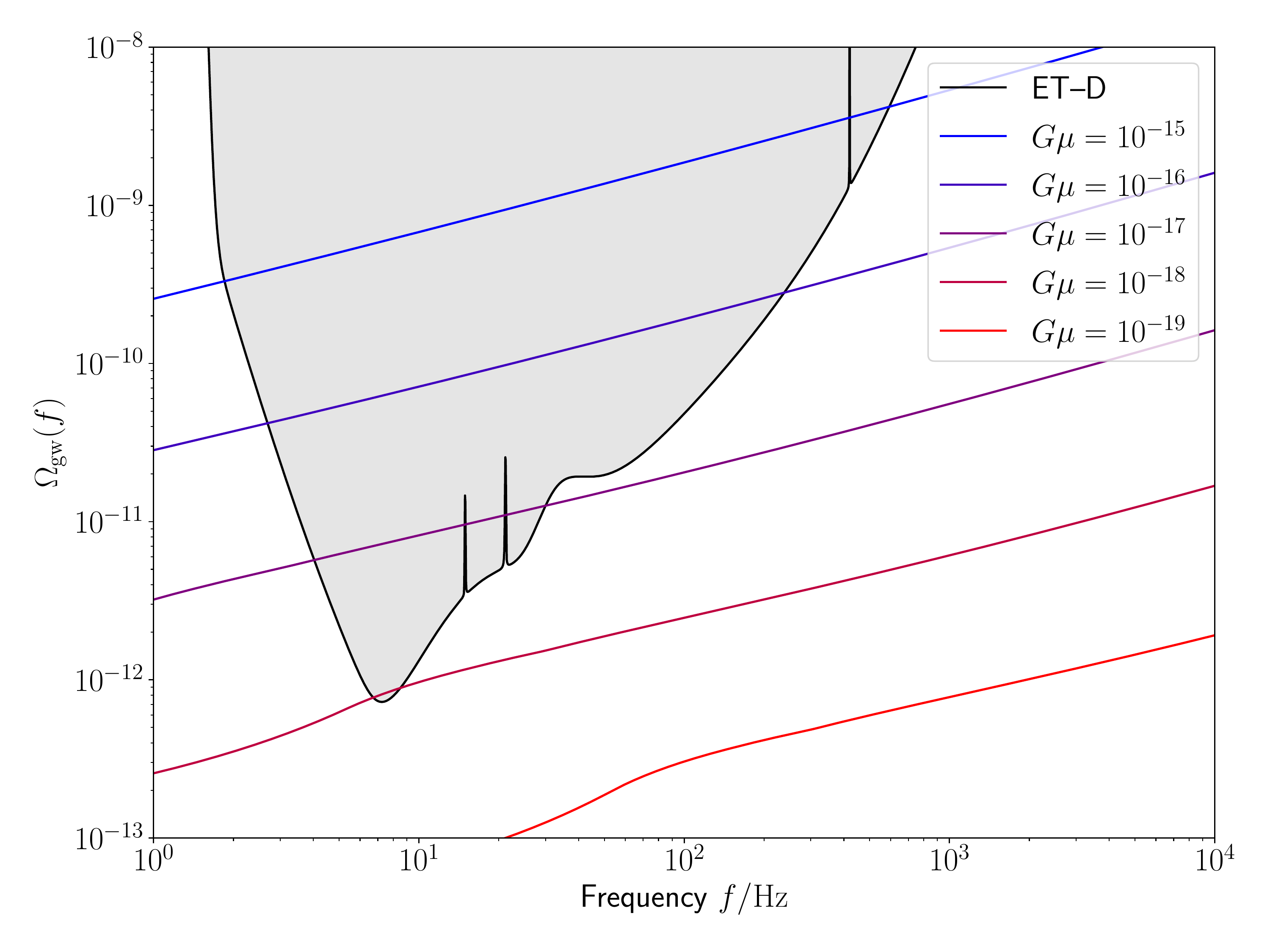}
\caption{The 95\% confidence detection region of ET-D for stochastic backgrounds  assuming  one year of observation time (shaded area), compared with the prediction for the stochastic GW background from cosmic strings, for  two different models of  the loop distribution discussed in ref.~\cite{Blanco-Pillado:2019nto} 
(left panel) 
and \cite{Lorenz:2010sm} 
(right panel). (Figure provided by A.C. Jenkins).  
\label{fig:sgwb-cs}}
\end{figure}

Phase transitions followed by spontaneous breaking of symmetries may leave behind topological defects as relics of the previous more symmetric phase of the Universe.
In the context of Grand Unified Theories, it has been shown \cite{Jeannerot:2003qv} that one-dimensional defects, called cosmic strings are generically formed. Cosmic string loops oscillate periodically in time, emitting GWs, which depend on a single parameter, the string tension $\mu$, related to the energy scale $\eta$ of the symmetry breaking through
\begin{equation}
G\mu\sim 10^{-6}\left(\frac{\eta}{10^{16}\, {\rm GeV} }  \right)^2~.
\end{equation}
Cosmic strings may emits bursts of beamed gravitational radiation.
The main sources of bursts are kinks, discontinuities on the tangent vector of a string resulting from string inter-commutations and exchange of partners, and cusps, points where the string instantaneously reaches the speed of light. Gravitational back reaction is expected to smooth out the string microstructure which implies that kinky loops become less wiggly and cusps may be the most important points for GW emission. Another mechanism leading to bursts of gravitational waves is kink-kink collisions, during which gravitational waves are emitted in all directions. The incoherent superposition of these bursts generates a stationary and almost Gaussian stochastic background of gravitational waves. Occasionally there may also be sharp and high-amplitude bursts of GWs above this stochastic background. 
A non-detection of a stochastic background of gravitational waves imposes bounds on the cosmic string tension and therefore on  particle physics models beyond the Standard Model.
ET  will be able to improve on 2G bounds by up to 8 orders of magnitudes. As we see from Fig.~\ref{fig:sgwb-cs},  with just one year of data one can detect or exclude values of $G\mu$ down to 
$10^{-17}-10^{-18}$, depending on the loop distribution.
Finally, within the Standard Model of particle physics, a stochastic background of GWs may be generated from a period of turbulence in the early Universe, which might arise for  the evolution of primordial magnetic fields coupled to the highly conducting magnetic fields ~\cite{Gogoberidze:2007an}.

The detection of a weak cosmological background requires the removal of the astrophysical foreground, which is a formidable challenge given the high expected number of signals from coalescing binaries. A simple subtraction of estimated waveforms will be insufficient since estimated waveforms do not perfectly match the true waveforms leaving sensitivity-limiting residuals in the data \cite{Regimbau:2017}. Consequently, advanced data-processing techniques need to be applied. The first method that was proposed makes it possible to project out the residuals \cite{Harms:2006}. A full Bayesian approach was suggested recently to achieve a combined estimation of the astrophysical and cosmological signals \cite{Smith:2018}. Both of these methods will push the limits of what will be computationally feasible. Therefore, the ultimate sensitivity achieved in searches of cosmological stochastic backgrounds might not only be limited by the instrument noise, but also by how effective data-analysis methods will be to understand the astrophysical foreground.

\subsubsection{Astrophysical backgrounds}

In addition to the cosmological background, an astrophysical contribution will  result from the superposition of a large number of unresolved sources too faint to be detected individually. Examples include short-lived burst sources, such as core collapses to neutron stars  or black holes, oscillation modes of (proto)-neutron stars, or the  final stage of compact binary mergers; periodic long lived sources, typically pulsars; or  the early inspiral phase of compact binaries or captures by supermassive black holes, whose frequency is expected to evolve very slowly compared to the observation time~\cite{Regimbau:2011rp}. The strongest astrophysical background in the frequency region of terrestrial detectors is expected to be due to the coalescence of binary black holes and binary neutron star systems.   To separate the cosmological and the astrophysical backgrounds the first step will be to use any distinct spectral dependence of the average (monopole) amplitude $\Omega_{\rm GW}$~\cite{Caprini:2019pxz}.   
However,  the better angular resolution of 3G detectors will most probably allow to spot the anisotropies (directionality dependence) of the astrophysical background. Such anisotropies contain information about the angular distribution of the sources and can be used as a tool for source separation as well as a tracer of astrophysical or cosmological structure. In fact the effects imprinted in the angular power spectrum of the stochastic background, due to the GW distorsion  by the intervening Large Scale Structure  distribution, like Kaiser, Doppler and gravitational potentials effects, can be used to study the Large Scale Structure and  make precision cosmology with 3G detectors \cite{Cusin:2017fwz, Cusin:2018rsq,Jenkins:2018uac,Jenkins:2018kxc,Jenkins:2019cau,Cusin:2019jpv,Bertacca:2019fnt,Canas-Herrera:2019npr}. 
For resolved sources, 3G detectors can join  astronomy facilities to enable thousands of host galaxy identifications from BNS and NSBH mergers. As discussed in sect.~\ref{sect:net3G}, for the best localized sources one could reach a localization accuracy below 1 deg$^2$ up to $z=0.5$, corresponding to multipoles $\ell \sim 100-1000$.
On the other hand, the resolution of 3G detectors for  the anisotropies of unresolved stochastic signals will be limited by the combination of the detector response and the baseline over which the cross-correlation of individual signals is being carried out (see \cite{Renzini:2018vkx} for a methodological study referring to 2G detectors).

Finally, thanks to the better sensitivity and angular resolution of 3G detectors, also alternative scenarios of production of GWs  will be tested:
 for instance, in U(1) extension of the Standard Model a spin-1 gauge boson, the dark photon, is predicted. If this particle is sufficiently light, it can produce an oscillating force on objects endowed with a dark charge which, on its turn, can bring to a stochastic GW signal potentially detectable by 3G detectors. Ultra-light boson clouds around spinning BH, that we already discussed in the context of continuous GW signals, can also produce a stochastic background due to the superposition of the signals produced by several decaying clouds.

\section{Summary of Key Science Questions}\label{sect:ScienceCaseKeyQuestions}

We conclude with a summary of the key scientific questions that ET will be able to tackle. As with any scientific enterprise of this scale, there will be questions for which, on the basis of our current understanding, we can say that ET is guaranteed to provide the answers, and questions that are more on the high risk/high gain side. We should not forget that ET will very much be a discovery machine: GW detection has literally opened a new window on the Universe. Thanks to third-generation detectors such as  ET we will now begin to look deeply  through this window. This means that we will  also penetrate  into uncharted territories, where further surprise could (and, in fact, likely, will) await for us.
A summary of the key science questions discussed above is as follows.

\vspace{1mm}

\begin{itemize}

\item ET will detect binary black hole coalescences up to redshift $z\sim 20$, over a broad mass range from sub-solar masses up to a few times $10^3 \Msun$, with a rate of order $10^5-10^6$ events per year. It  will therefore provide a census of the population of BHs across the whole epoch of star formation and beyond, answering crucial questions on the progenitors, formation, binary evolution and demographics of stellar BHs. The astrophysical discovery potential in this direction is guaranteed. A detector network would of course result in further benefits,\footnote{As we have discussed in sect.~\ref{sect:net3G}, apart from reducing the false alarm rate and increasing the overall signal-to-noise ratio, a detector network is in general important  for accurate source localization, which also gives  the possibility of identifying an electromagnetic counterpart. For BH-BH binaries the latter point is less relevant, since anyhow BH binaries are not expected to have an electromagnetic counterpart. However,  as we discussed, accurate localization could  provide  a characterization of the anisotropies of the binary BH population, allowing  for instance  to perform studies of correlation with large-scale structure surveys.}  but even  ET as a single detector is fully adequate to address these issues. 

\vspace{1mm}

\item ET will extend  the region of BH masses  explored by 2G detectors, which is limited to  $O(100) \Msun$, to the range from several hundreds solar masses (that could be detected up to redshifts of order 10 or more, see the left panel in Fig.~\ref{fig:gw_horizons}) to several thousands solar masses (that could be detected up to $z\sim 1-5$).  This 
opens the possibility of detecting  these intermediate mass BHs, providing the first clear evidence for their existence and  studying the possibility that they are the seeds of the supermassive BHs in the center of galaxies. The detections  could be performed by a single 3G detectors such as ET, and would  complement in an ideal manner the observations expected from  LISA: for masses $M$ in the range $(10^2-10^4) \Msun$ LISA would observe the long inspiral phase, that could stay in its bandwidth for months
or even years, while a 3G detector could see the final and much louder merger phase of the same events.

\vspace{1mm}

\item ET will detect the coalescence of binary neutron stars up to $z\simeq 2-3$, with a rate of order $10^5$ events per year. This range  reaches the peak of the star formation rate and therefore covers the vast majority of NS binaries  coalescing throughout the Universe. This will allow us to investigate their formation mechanisms, evolution and demographics. The sensitivity of ET in the high-frequency regime will allow us to access the GW signal of the merger phase, that is unaccessible to 2G detectors and carries detailed information on the internal structure of neutron stars and on their equation of state. This will have  important implications also for fundamental physics, allowing us to study QCD at ultra-high density and  the possibility of phase transitions in the NS core, such as a transition to deconfined quarks or the formation of exotic states of matter. These detections, and a rich science output coming from them,  are guaranteed. Again, these goals can be obtained even by ET as a single detector. A network of three 3G detectors would bring, on top of this, the possibility of accurate localization of the source, allowing to give to electromagnetic telescope the information necessary to identify an electromagnetic counterpart and perform multi-messenger studies.

\vspace{1mm}

\item ET  could detect several new astrophysical sources of GWs, such as signals emitted during core collapse supernovae, continuous signals from isolated rotating NSs, and possibly burst signals from NSs. While not guaranteed, these signals would bring rich information. Detecting  the GWs from supernovae would
elucidate the mechanisms of supernova explosions and its post-collapse phase.
The detection of continuous GWs from NSs  would   allow us to explore the condition of formation and evolutions  of isolated NS, providing information on  their spin, thermal evolution and magnetic field.  ET will be able to detect `mountains' on the surface of a NS as small as $10^{-2}$~mm,   which in turn would again give us  information on the inner structure of NS and on the corresponding aspects of nuclear and particle physics, such as the existence of exotic matter in the NS core. These goals can be obtained by ET as a single detector. Furthermore, in the case of rotating NSs, a single detector would accurately localize the source, thanks to the Doppler effects from the Earth movement during the observation of the signal, that can go on for months.

\vspace{1mm}

\item The waveform from the loudest BH-BH and NS-NS coalescences will be observed by 
ET with exquisite precision. This will allow accurate tests of General Relativity, both in the inspiral phase, where one can test the validity of the post-Newtonian expansion of GR to sub-permille accuracy, and in the merger and post-merger phase. The latter is particularly interesting since it would allow us to test the nature of BHs and the dynamics of space-time close to the horizon of the final BH, through the observation of  the frequencies and lifetimes of its longer-lived quasi-normal modes. This would allow us to perform for the first time accurate quantitative tests of the predictions of GR in this extreme domain. The possibility of performing such accurate tests is guaranteed, and can also  be performed by ET as a single detector. These tests could also in principle lead to  surprises, such as revealing  the existence of exotic compact objects, and could even carry observable imprints of quantum gravity effects. While the latter goals are more speculative, their impact would be revolutionary.

\vspace{1mm}

\item ET will test several dark matter candidates. If dark matter is made, at least in part, by primordial BHs in the mass range $\sim 0.1 - 100\,M_\odot$, ET will be able to provide definite evidence for them. Indeed, thanks to its extraordinary reach, ET could observe BH binaries at redshift  $z\sim 10-20$, before
any stars had formed that could create black holes in the usual way; even a single event observed at such redshift would necessary have a non-stellar origin. Furthermore, thanks to the extension of its frequency band toward both  low and high frequencies, ET will detect  BHs across a large spectrum of masses. The detection of even a  single BH of sub-solar mass would again point clearly to a non-stellar origin. This topic belongs to those whose success is not guaranteed (observed BHs could, after all, have just stellar origin) but certainly belongs to the high-gain category. Showing that at least a fraction of the observed BHs are of primordial origin would be a discovery of fundamental importance not only in astrophysics but also from the point of view of fundamental physics and cosmology, providing unique information on primordial inflation and on physics at correspondingly high energies. Another dark matter candidate that will be tested by ET is an ultralight boson such as a  light axion, that could form a bosonic cloud near a BH,
or more generally dark matter particles that are captured by compact objects and accumulate in the core of a NS, or that accumulate near compact objects and create a drag in the dynamics of a compact binary.
 ET will be able to explore these possibilities even as a single detector.  

\vspace{1mm}

\item ET will explore the nature of dark energy and the possibility of modifications of General Relativity at cosmological distances. The crucial point here is again the ability to detect compact binary coalescences up to cosmological distances, providing an absolute measurement of their distance. The relation between luminosity distance and redshift, in the range of redshifts explored by  ET, carries very distinctive signature of the dark energy sector of a modified gravity theory, through the dark energy equation of state and, especially, through  an observable related to modified GW propagation. The latter is a particularly powerful probe of dark energy, which is accessible only to GW experiments. Thanks to the large number of detections, ET could probe the luminosity-distance relation even as a single detector using statistical techniques, but this topic would significantly benefit from the presence of a network of GW detectors leading to accurate localization of many sources (allowing for  the measurement of the redshift from electromagnetic follow-up observations, as with GW170817), or from synergies with gamma-ray burst detectors. From the point of view of cosmology, ET is guaranteed to obtain  important results (accurate measurement of $H_0$, significant limits on the equation of state of dark energy), complementary to measurements obtained with electromagnetic probes. The possibility of detecting  modifications of General Relativity at cosmological scales and understanding the origin of dark energy is not guaranteed, but would be revolutionary.

\vspace{1mm}

\item ET will search for stochastic backgrounds of GWs, which are relics of the earliest cosmological epochs. 
Such backgrounds, if detected, would carry information of the earliest moment of the Universe, and on physics at the corresponding high-energy scales, that would be unaccessible  by electromagnetic (or neutrino) observations. ET, thanks to its design corresponding to three independent interferometers, could obtain significant limits on stochastic backgrounds already as a single detector, although  a network of two well-separated 3G detectors  would have a better rejection of local background noise and would allow to resolve the angular anisotropies of the background.
In particular,  ET can contribute to shedding light on early universe models of inflation and is also sensitive to features that will allow the disentangle a cosmological background from an astrophysical one. It can test the post-inflationary period through GW from a (p)reheating stage after inflation and from a stiff phase of evolution. It  will have enough angular resolution to distinguish galactic from extragalactic backgrounds of GW through the characterization and mapping of the angular distribution of GW anisotropies both from astrophysical and cosmological sources.  
Stochastic backgrounds of cosmological origin in the ET frequency window depend on physics beyond the Standard Model. Thus, the  predictions are unavoidably uncertain, and the gain from a successful detection would be correspondingly high, allowing us to explore the earliest moments after the big bang.

\end{itemize}

\vspace{5mm}\noindent
{\bf Acknowledgments}.
We thank Michele Punturo  for very useful feedback, and the referee for very useful suggestions.
The work  of EB, SF and MM is supported by the  Swiss National Science Foundation and  by the SwissMap National Center for Competence in Research. NB, DB and SM acknowledge partial financial support by ASI Grant No. 2016-24-H.0.
JGB acknowledges support from the Research Project PGC2018-094773-B-C32 (MINECO-FEDER) and the Centro de Excelencia Severo Ochoa Program SEV-2016-0597.   TH is supported by the DeltaITP and NWO Projectruimte grant GW-EM NS.
The work of CP is supported by the  Istituto Nazionale di Fisica Nucleare (INFN). MS is supported in part by the Science and Technology Facility Council (STFC), United Kingdom, under the Research Grant ST/P000258/1.

\bibliographystyle{utphys}
\bibliography{refsScienceCase}

\end{document}